\documentclass[journal]{IEEEtran}
\usepackage{booktabs}
\usepackage{tabularx}
\usepackage{amsmath,amssymb}
\usepackage{graphicx}
\usepackage{epstopdf}
\usepackage{float}
\usepackage{multirow}
\usepackage{cite}
\usepackage{booktabs}
\usepackage{subfigure}
\usepackage{xcolor}
\usepackage{amsfonts}
\usepackage{mathrsfs}
\usepackage{pifont}
\usepackage{stfloats}
\usepackage{color}
\usepackage{booktabs}
\usepackage{caption}
\usepackage{algorithm}
\usepackage{caption}
\usepackage{algorithmicx}
\usepackage{algpseudocode}
\renewcommand{\t}{^{\mbox{\tiny {T}}}}
\newcommand{\diag}{\mbox{diag}}
\IEEEoverridecommandlockouts                              
\newcommand{\m}{{\rm m}}

\newcommand{\eproof}{\hfill\rule{2mm}{2mm}}

\newcommand{\bstate}{\begin{state} }
	\newcommand{\estate}{ \hfill  \rule{1mm}{2mm}\end{state}}

\newcommand{\bass}{\begin{ass} }
	\newcommand{\eass}{ \hfill  \rule{1mm}{2mm}\end{ass}}

\newcommand{\brem}{ \begin{remark}  }
	\newcommand{\erem}{\hfill \rule{1mm}{2mm}
\end{remark} }
\newcommand{\bthm}{\begin{theorem}  }
	\newcommand{\ethm}{ \hfill  \rule{1mm}{2mm}
\end{theorem} }
\newcommand{\blem}{\begin{lemma}  }
	\newcommand{\elem}{ \hfill \rule{1mm}{2mm}
\end{lemma} }
\newcommand{\bcorollary}{\begin{corollary}  }
	\newcommand{\ecorollary}{  \hfill \rule{1mm}{2mm}
\end{corollary} }
\newcommand{\bdefn}{\begin{definition}}
	\newcommand{\edefn}{  \hfill \rule{1mm}{2mm}
\end{definition} }
\newcommand{\bproposition}{\begin{proposition} }
	\newcommand{\eproposition}{\hfill \rule{1mm}{2mm}
\end{proposition} }
\newcommand{\bexample}{\begin{example} \rm}
	\newcommand{\eexample}{ \hfill \rule{1mm}{2mm}
\end{example} }

\newcommand{\bcon}{\begin{condition} \rm}
	\newcommand{\econ}{ \hfill \rule{1mm}{2mm}
\end{condition} }

\newcommand{\proofnow}{\noindent{\bf Proof: }}

\newtheorem{theorem}{\bf Theorem}[section]
\newtheorem{ass}{\bf Assumption}[section]
\newtheorem{lemma}{\bf Lemma}[section]
\newtheorem{definition}{\bf Definition}[section]
\newtheorem{remark}{\bf Remark}[section]
\newtheorem{corollary}{\bf Corollary}[section]
\newtheorem{proposition}{\bf Proposition}[section]
\newtheorem{example}{\bf Example}[section]
\newtheorem{condition}{\bf Condition}[section]
\newtheorem{state}{\bf Assumption}[section]

\newcommand{\sint}{\textstyle{\int}}

\renewcommand{\Re}{\operatorname{Re}}
\renewcommand{\Im}{\operatorname{Im}}

\newcommand{\blue}{\color{black}}
\DeclareMathOperator{\rank}{rank}
\usepackage{fancyhdr}
\allowdisplaybreaks
\begin{document}
\title{Prescribed-time Cooperative Output Regulation of  Linear Heterogeneous Multi-agent Systems}
\author{Gewei Zuo, Lijun Zhu, Yujuan Wang and Zhiyong Chen
\thanks{G. Zuo and L. Zhu are with School of Artificial Intelligence and Automation, Huazhong University of Science and Technology, Wuhan 430072, China. L. Zhu is also with Key Laboratory of Imaging Processing and Intelligence Control, Huazhong University of Science and Technology, Wuhan 430074, China (Emails: gwzuo@hust.edu.cn;  ljzhu@hust.edu.cn).
	}
\thanks { Y. Wang is with the State Key Laboratory of Power Transmission Equipment \& System Security and New Technology, and School of Automation, Chongqing University, Chongqing, 400044, China (Email: yjwang66@cqu.edu.cn).}
\thanks {Z. Chen is with School of Engineering, The University of Newcastle, Callaghan, NSW 2308, Australia (Email: chen@newcastle. edu.cn). }}

\markboth{}{ \MakeLowercase{\textit{et al.}}: }

\maketitle

%
%


{\blue
\begin{abstract}
This paper investigates the prescribed-time cooperative output regulation (PTCOR) for a class of linear heterogeneous multi-agent systems (MASs) under a directed communication graph.
As a special case of PTCOR, the necessary and sufficient condition
for prescribed-time output regulation of an individual system is
first explored, while only sufficient condition is discussed in the literature.
%
A PTCOR algorithm is subsequently developed, which is composed of prescribed-time distributed observers, local state observers, and tracking controllers, utilizing a distributed feedforward method.
This approach converts the PTCOR problem into
the prescribed-time stabilization problem of a  cascaded subsystem.  The criterion for the prescribed-time stabilization of the cascaded system is proposed, which differs from that of traditional asymptotic or finite-time  stabilization of a cascaded system.
It is proved that the regulated outputs converge to zero within a prescribed time and remain as zero afterwards, while all internal signals in the closed-loop MASs are uniformly bound. Finally, the theoretical results are validated through two numerical examples.
\end{abstract}
}


{\it Copyright Declaration}:
This work has been submitted to the IEEE for possible publication. Copyright may be transferred without notice, after which this version may no longer be accessible.

\IEEEpeerreviewmaketitle
\section{Introduction}
{\blue
In recent decades, the cooperative output regulation (COR) problem for MASs has attracted considerable research attention, owing to its wide-ranging applications, such as flight formation control \cite{ABDESSAMEUD50}, multi-vehicle coordination \cite{ren2007distributed51}, and power balance in microgrids \cite{Caihe52}. The COR problem aims to  design a distributed controller for each agent to track the reference input  generated by an exosystem that acts as the leader within the MASs.  This problem extends the classical output regulation problem \cite{huang3} from a single system to the context of MASs, where the state of the exosystem is only accessible to a subset of agents.
There are two primary approaches to solving the COR problem: the distributed feedforward method \cite{yaghmaie2016output55,abdessameud2018distributed15,seyboth2016cooperative56,kawamura2020distributed57} and the distributed internal model method \cite{deutscher2021robust63,wieland2011internal58,koru2020cooperative64}. However, these results in \cite{yaghmaie2016output55,abdessameud2018distributed15,seyboth2016cooperative56,kawamura2020distributed57,deutscher2021robust63,wieland2011internal58,koru2020cooperative64} primarily focus on the asymptotic stability, where the regulated outputs converge to zero as time approaches infinity. The asymptotic convergence approach, although effective in certain scenarios, may fail to meet the convergence time requirements.
In contrast,   finite-time control (FTC) originally proposed in \cite{weiss1965stability100,weiss1967finite101}, offers improved  convergence performance and robustness. FTC has since been applied to the control of various types of MASs \cite{franceschelli2014finite53, ghasemi2014finite54,zhao2016distributed40, sarrafan2021bounded59, huang2015adaptive38, li2018finite41}. In the context of COR, the finite-time approach, employing fractional-power feedback and distributed observers, is explored in \cite{wu2021finite20,wu2023finite102}, where the settling time of regulated outputs is bounded. However, this settling time depends on initial conditions and design parameters, and can be accurately estimated only when the initial conditions are known \cite{zhao2016distributed40}.
Moreover, due to varying initial conditions and design parameters, the settling times differ across agents in the MASs.
To address this issue, fixed-time COR has been introduced for linear heterogeneous MASs \cite{song2021distributed103,zhang2022cooperative104}. In this approach, fractional-power feedback and feedback with powers greater than one ensure the fixed-time convergence, where the settling time is independent of initial conditions. Although the   convergence  time is fixed and  uniform among agents, the settling time is still affected by control parameters and   it cannot be specified \emph{a priori}.

Prescribed-time control (PTC) is subsequently proposed in \cite{wyj1}, where a class of time-varying feedback gains, which increase to infinity as the system approaches the prescribed time, are introduced into the feedback loop. This approach offers a distinct advantage: the settling time can be specified \emph{a priori}, and it remains independent of initial conditions and any controller parameters.  Following \cite{wyj1}, PTC has been extended to a broader range of MASs, as seen in \cite{wang2018leader44,ning2020bipartite42, ning2018prescribed43,wangcontainment48,cui2023sliding81}. Additionally, in \cite{chen2023prescribed97}, the PTCOR problem for the linear heterogeneous MASs is addressed, where the prescribed-time distributed observers are developed.


\begin{table*}[ht]
\centering
\caption{{\blue Comparisons on the Existing Results of COR} }
\vspace{3mm}
\label{table1}
\begin{tabular}{*{6}{>{\centering\arraybackslash}p{2.5cm}}}
\toprule[1pt]
\textbf{Items} & \textbf{Convergence Speed} & \multicolumn{2}{c}{\textbf{Settling-Time}}  & \multicolumn{2}{c}{\textbf{State-of-Art}} \\
               &                             & {Initial Conditions Free} & {Control Parameters Free} & Sufficient and Necessary Condition &  Control Criterion  \\ \toprule[1pt]
\cite{yaghmaie2016output55,abdessameud2018distributed15, seyboth2016cooperative56,kawamura2020distributed57, deutscher2021robust63,wieland2011internal58,koru2020cooperative64}             & Asymptotic                           & ---                   &--- &Sufficient Condition&---                     \\ \toprule[0.5pt]
\cite{wu2021finite20,wu2023finite102}              & Finite-time &---&---&Sufficient Condition&---                      \\ \toprule[0.5pt]
\cite{song2021distributed103,zhang2022cooperative104}& Fixed-time& \checkmark&---&Sufficient Condition & ---                      \\ \toprule[0.5pt]
 \cite{chen2023prescribed97}  &Prescribed-time &\checkmark & \checkmark & Sufficient Condition & ---\\\toprule[0.5pt]
Proposed Algorithm & Prescribed-time & \checkmark &\checkmark & Sufficient and Necessary Condition & A Criterion for Prescribed-Time Stabilization of Cascaded Systems \\ \toprule[1pt]
\end{tabular}
\end{table*}

This paper further explores the PTCOR for linear heterogeneous MASs. The work \cite{chen2023prescribed97} focus solely on sufficient conditions for the PTCOR, while we   establish  both necessary and sufficient conditions for a special case of PTCOR in a class of typical linear heterogeneous MASs under state feedback and output measurement feedback.
The comparisons between the proposed scheme and the existing methods are shown in Table. \ref{table1}.
The main contributions and novelties of our approach are outlined as follows:

(1) We derive the necessary and sufficient conditions for the solvability of prescribed-time output regulation (PTOR) for an individual system, which is a prerequisite to ensuring the prescribed-time convergence of both the distributed observers and the closed-loop system for the PTCOR. 
The work  \cite{chen2023prescribed97} proposes the
 sufficient condition for the PTCOR   based on a set of Linear Matrix Inequalities (LMIs). Given the necessary condition for the PTCOR, we  derive  direct and concise algebraic conditions for the  PTCOR.  

(2) By utilizing the distributed feedforward method, the PTCOR problem is transformed into a prescribed-time stabilization problem involving local tracking errors, distributed estimate errors, and local estimate errors. The subsystems composed of distributed and local estimate errors, and that of local tracking errors, form a cascaded system, where the state of the first subsystem acts as the input to the second subsystem.
A novel criterion for the  prescribed-time stabilization of  the cascaded system is  proposed. It is observed that achieving the prescribed-time stabilization in cascaded systems requires more stringent conditions than those necessary for the asymptotic or finite-time stabilization. In particular, the   controller gain design for the first subsystem must also consider the effect of the second subsystem.

(3) To the best of our knowledge, the proposed criterion for the prescribed-time stabilization of a cascaded system not only ensures the prescribed-time convergence in the PTCOR problem but also generalizes and strengthens the results presented in \cite{wang2018leader44, ning2020bipartite42, ning2018prescribed43, wangcontainment48, cui2023sliding81}.
By choosing suitable parameters, the closed-loop systems satisfy the criterion of prescribed-time convergence, and thus all the regulated outputs converge to zero within prescribed-time and remain as zero afterwards. Furthermore, the internal signals in the closed-loop MASs are proved to be uniformly bounded over infinite time interval.
}

 {\blue The      paper is structured as follows. Section \ref{sec2} describes the problem formulation. Section \ref{sec3} identifies the sufficient and necessary condition for the implementation of PTCOR. In Section \ref{CasSys},  we discuss how the  PTCOR problem is converted into a prescribed-time stabilization problem of a cascaded system and introduce a criterion of prescribed-time convergence for the cascaded system. Section \ref{sec:StaAna} focuses on  the stability analysis and implementation of PTCOR. The numerical simulation is conducted in Section \label{Sec:Simulation} and the paper is concluded in Section \ref{sec:conclusion}.   }
%

{\it Notations:}
$\mathbb R$, $\mathbb R_{\geq 0}$ and $\mathbb R^n$ denote the set of real numbers, the set of non-negative real numbers, and the $n$-dimensional Euclidean space, respectively.
The set of eigenvalues of a square matrix $A$ is denoted as $\lambda(A)$. If the elements of $\lambda(A)$ are all real numbers, $\lambda_{\min}(A)=\min(\lambda(A))$ and $\lambda_{\max}(A)=\max(\lambda(A))$. For $x\in\mathbb{R}^n$ and $C\in\mathbb{R}^{n\times n}$,  $\| Cx\|\leq \| C\| \| x\|$, where $\|x\|=x\t x$ denotes the Euclidean norm and $\| C\|$ is any norm compatible with the Euclidean norm of $n$-dimensional vector. {\blue The symbol $1_N \in \mathbb R^N$ (or $0_N \in \mathbb R^N$) denotes an $N$-dimensional column vector whose elements are all $1$ (or $0$), and $I_N$ denotes the $N$-dimensional identity matrix.  The symbol $\otimes$ represents Kronecker product. }

\section{Problem Formulation}\label{sec2}
Consider the linear MASs as follows
\begin{equation}\label{sys1}
	\begin{aligned}
			\dot{x}_i&=A_ix_i+B_iu_i+E_i\upsilon_0 \\
		e_i&=C_ix_i+D_iu_i+F_i\upsilon_0\\
		y_i &=C^\m_i  x_i+D^\m_i  u_i+F^\m_i  \upsilon_0,\; i=1,\cdots,N
		\end{aligned}
	\end{equation}
where $x_i\in \mathbb{R}^{n_i}$, $u_i\in \mathbb{R}^{m_i}$, $e_i\in\mathbb{R}^{p_i}$, and $y_i  \in\mathbb{R}^{p^\m_i  }$ are the state, control input, regulated output, and measurement output of the $i$-th subsystem, respectively.
The exogenous signal $\upsilon_0\in\mathbb{R}^{q}$ represents
the reference input to be tracked and it  is assumed to be generated by the  exosystem
\begin{equation}\label{sys2}
	\dot{\upsilon}_0=S_0\upsilon_0
\end{equation}
for a matrix $S_0\in\mathbb{R}^{q\times q}$.  Exosystem (\ref{sys2}) exhibits  neutral stability, i.e., the eigenvalues of matrix $S_0$ are all in the left closed plane, and the eigenvalues with zero real part are semi-simple.

We associate the node $0$ with the  system (\ref{sys2}) and call it a leader,
while the $N$ agents in (\ref{sys1}) are called followers.
Let $\mathcal{G}=(\mathcal{V},\mathcal{E})$ denote the directed graph associated with this leader-following network, where the node set is $\mathcal{V}=\left\{0,1,\cdots,N\right\}$ and the edge set is $\mathcal{E}\subseteq\mathcal{V}\times\mathcal{V}$.
Each edge $(i,j)\in\mathcal{E}$ symbolizes the transmission of information from agent $i$ to agent $j$. Agent $i$ is deemed a neighbor of agent $j$ if the edge $(i,j)\in \mathcal{E}$.
Denote
the node set $\mathcal{\bar V}=\left\{1,\cdots,N\right\}$ excluding node $0$.
{\blue Denote by $\mathcal{A}=[a_{ij}]\in \mathbb R^{(N+1)\times (N+1)}$  the weighted adjacency matrix of $\mathcal G$, where $(j,i)\in \mathcal E \Leftrightarrow a_{ij}>0$, and $a_{ij}=0$ otherwise. A self edge is not allowed, i.e., $a_{ii}=0$. The Laplacian matrix of the graph is  denoted as $\mathcal{L}=[l_{ij}]\in \mathbb{R}^{(N+1)\times(N+1)}$,
where $l_{ii}=\sum_{j=0}^{N}a_{ij}$, $l_{ij}=-a_{ij}$ with $i\neq j$.}

 {\blue  This paper explores a scenario where the signal to be tracked may affects the system's dynamics, with only a portion of the nodes having access to the state of exosystem (\ref{sys2}). }  For example,
in an autonomous underwater vehicle (AUV) fleet for ocean monitoring, it is a typical task for vehicles to follow the reference trajectories of a leading vehicle. As the AUVs share the same underwater environment, their physical coupling between the leader and the followers arises from the fact that the motion of the leader changes the environment and hence affects the agents in its proximity.   The  COR problem of  (\ref{sys1}) in the sense of asymptotical convergence has been studied, for instance, in \cite{yaghmaie2016output55,abdessameud2018distributed15,seyboth2016cooperative56,kawamura2020distributed57}.

%

This paper investigates the PTCOP problem
for the MASs \eqref{sys1}. First, we give the rigorous definition
of  prescribed-time convergence for a dynamic system. 

{\blue
\bdefn \cite[Definition 4.3]{2002Nonlinear99} \label{def:1}
A   continuous function  $\beta:[0,c)\times [0,\infty )\mapsto [0,\infty)$ is said to belong to class $\mathcal {KL}_T$  if  for each fixed $s$, the mapping $\beta(r,s)$ belongs to class $\mathcal K$ with respect to $r$, where class $\mathcal K$ is defined in Definition 4.2 of \cite{2002Nonlinear99}. Additionally, for each fixed $r$, there exists a constant $T$ such that, for $s\in [0,T)$, the mapping $\beta (r,s)$ is decreasing with respect to $s$ and satisfies $\beta(r,s)\to 0$ as $s\to T$, $\beta(r,s)=0$ for $s\in [T,\infty)$.
\edefn
}

{\blue
\bdefn \cite{wyj1,holloway2019prescribed106}\label{def:ptfc}
Consider the system
\begin{equation}\label{chisys-1}
\begin{aligned}
\dot\chi&=f(t,\chi),\quad \chi(t_0)=\chi_0\\
e&=h(t,\chi)
\end{aligned}
\end{equation}
where $\chi \in\mathbb R^n$ represents the state, $e\in\mathbb R^p$ is the output, and $\chi_0$ denotes the initial state at $t=t_0$. The system output is said to achieve the prescribed-time convergence (towards zero within $T+t_0$ and remains as zero afterwards) if there exists a predesigned time  $T$ along with a corresponding  function $\beta\in\mathcal{KL}_T$ such that, for any $\chi_0$,
\begin{equation}
	\| e(t)\| \leq \beta( \|\chi_0\|,t-t_0) \label{ptc}
\end{equation}
holds for $t\geq 0$.
\edefn
}

\brem
{\blue
In Definition \ref{def:1}, the class of $\mathcal {KL}_T$ functions is defined as an extension to $\mathcal {KL}$  (Definition 4.3 in \cite{2002Nonlinear99}) functions  with different domains, which is used for the  analysis of prescribed-time stability.  }
In the above definition, the settling time $T$ of the prescribed-time convergence is independent of
the initial state $\chi_0$. In the literature, the weaker requirement that $T$ depends on the initial state $\chi_0$
is called finite-time convergence \cite{wu2021finite20,wu2023finite102}. In the so-called fixed-time convergence, $T$ is also independent of $\chi_0$,
but only the existence of $T$ is guaranteed \cite{song2021distributed103,zhang2022cooperative104}. In other words, $T$ cannot be specified \emph{a priori} in the fixed-time convergence.
\erem

Define a piecewise continuous function
\begin{equation}\label{mu}
	\mu (t)=\left\{\begin{array}{cc}
		\frac{1}{T+t_0-t},& t\in[t_0,T+t_0)\\
		a ,& t\in[T+t_0,\infty)
		\end{array}\right .
	\end{equation}
where $T,a>0$. For simplicity, $\mu (t)$ is denoted as $\mu$ if no confusion occurs. Without losing generality,  we can set $a=1/T$, ensuring that $\mu^{-1}(t)\leq T$ for $t\geq t_0$.

Given that the state $\upsilon_0$ of the leader (\ref{sys2}) is solely available to the connected agents in the graph rather than all followers, it's necessary to construct a distributed observer for each follower to acquire an estimate of the leader's state $\upsilon_0$,
\begin{equation}
	\dot{\upsilon}_i=S_0\upsilon_i+\psi\mu(t)\sum_{j=0}^Na_{ij}(\upsilon_j-\upsilon_i),\;  \forall t\geq t_0, i\in\mathcal{\bar V} \label{18}
\end{equation}
{\blue where $\psi>0$ is a design parameter and $a_{ij}$ is the element of adjacency matrix $\mathcal A$. }


 By utilizing the estimated state $\upsilon_i$ in (\ref{18}), we aim to propose two feedback controllers.
The  first one is the state feedback controller, for  $i\in\mathcal{\bar V}$, designed as
\begin{equation}
	u_i=\bar K_i x_i+\tilde K_i \upsilon_i+\mu(t) K_i(x_i-X_{i}\upsilon_i). \label{eq11}
\end{equation}
This scenario can be viewed as a special case where the measurement output $y_i=x_i$.
The second one is the measurement output feedback controller, for $ i\in\mathcal{\bar V}$,
\begin{align}
	\dot{\hat{x}}_i&=A_i\hat{x}_i+B_iu_i+E_i\upsilon_i+(L_{i}+\mu(t) \tilde L_{i})   \notag\\
	&\quad   \times (y_i-C^\m_i  \hat{x}_i-D^\m_i  u_i-F^\m_i  \upsilon_i)\label{hatx}  \\
	u_i&=\bar K_i\hat{x}_i+\tilde K_i\upsilon_i+\mu(t) K_i(\hat{x}_i-X_{i}\upsilon_i).
	\label{40}
\end{align}
{\blue where $\times$ represents the multiplication operation.}

The  parameters $X_i$, $\bar K_i$,  $\tilde K_i$,  $K_i$, $L_{i}$, and $\tilde L_{i}$ in the two controllers are to be designed.
Note that $\hat x_i$-dynamic
is called the local state observer.
The controller without the
terms associated with $\mu(t)$ can achieve traditional COR \cite{huangjie1} but these additional $\mu(t)$-dependent terms are introduced to ensure prescribed-time  convergence.

%
 Note that the time-varying term $\mu(t)$ used in the design is unbounded as $\lim_{t\rightarrow T+t_0}\mu(t) =\infty$.
However, the proposed design must guarantee that the $\mu(t)$-dependent terms in  (\ref{18}),  (\ref{eq11}), \eqref{hatx}, and \eqref{40}, denoted as
\begin{equation}\label {eq:phi_signal}
\begin{aligned}
	\phi_1(t) &= \mu(t) (\upsilon_j-\upsilon_i)  \\
	\phi_2(t) &= \mu(t)(x_i-X_{i}\upsilon_i) \\
	\phi_3(t) &=\mu(t)  (y_i-C^\m_i  \hat{x}_i-D^\m_i  u_i-F^\m_i  \upsilon_i) \\
	\phi_4(t) &=\mu(t) (\hat{x}_i-X_{i}\upsilon_i)
\end{aligned}
\end{equation}
are bounded for all $t\geq t_0$, which facilitates implementation of the controllers. To simplify mathematic derivations, we define two types of Lyapunov functions for the prescribed-time stabilization.

	\bdefn\label{def:PTISSLF}
	Consider the system $\dot x=f(t,x,z)$ with $x\in\mathbb{R}^n$ and $z\in\mathbb{R}^q$.
	The continuous differential function $V(x):\mathbb R^n\mapsto \mathbb R_{\geq 0}$ is called the prescribed-time input-to-state stable Lyapunov function (PTISSLF)
	 for the system if $V(x)$ and its derivative along the trajectory of the system satisfy, for all $x \in \mathbb R^n$,
	\begin{equation}\label{PTISSLF}
		\begin{gathered}
			\underline \alpha\|x\|^2\leq V(x)\leq \bar \alpha\|x\|^2\\
			\dot V(x)\leq -\alpha \mu V(x)+\tilde \alpha V(x)+{\blue \sigma\mu^m \|z\|^p }
		\end{gathered}
	\end{equation}
	where $\underline \alpha$, $\bar \alpha$, $\alpha$, $\tilde \alpha$, $\sigma$, $m$ and {\blue $p$}  are positive finite constants.
When $\sigma=0$, the continuous differential function $V(x)$ is called the prescribed-time   Lyapunov function (PTLF)
 for the system if $V(x)$ and its derivative along the trajectory of the system satisfies (\ref{PTISSLF}) without the term {\blue $\sigma\mu^m(t) \|z\|^{p}$}.
	\edefn

{\blue
\brem
The PTISSLF in \eqref{PTISSLF} is different from the  input-to-state stable Laypunov function in \cite{panteley2001growth94, nguyen2024formation96} and finite-time or fixed-time input-to-state stable Lyapunov function \cite{ding2012nonsmooth93,wang2024fixed95}. The difference lies in that  the term $\tilde \alpha V(x)$ is allowed   even if it   causes a divergent term in the boundedness result of $V(x)$.  Furthermore, the PTISSLF in \eqref{PTISSLF} is a generalization to the definitions given in  \cite{wyj1,hua2021adaptive69,ZuoGeWei70,wang2018leader44,ning2023event86}.
\erem
}

\section{Sufficient and Necessary Condition}\label{sec3}
In this section, we will initially examine the straightforward scenario where there is only one follower, i.e., $N=1$. In such a scenario, the PTCOR problem simplifies to the PTOR problem.
We discuss a  sufficient and necessary condition for the PTOR problem  when the state feedback law (\ref{eq11}) or measurement output feedback law (\ref{hatx}), (\ref{40}) is used.
The condition is also required for PTCOR to ensure  prescribed-time convergence of
the distributed observers as well as the closed-loop MASs.

The following two technical assumptions are commonly applied for the COR problem.
\bass \label{ass3}
For any $c\in\lambda(S_0)$,
\begin{equation*}\label{equa18}
	\rank\left [ \begin{array}{cc}A_i-c I&B_i\\C_i&D_i\end{array}\right ]=n_i+p_i, \quad \forall i\in\mathcal{\bar V}
\end{equation*}
{\blue where $n_i$ and $p_i$ are the dimensions of state $x_i$ and regulated output $e_i$ of $i$-th agent, respectively.}
\eass
\bass \label{ass4}
The graph $\mathcal{G}$ contains a spanning tree with the node $0$ as the root.
\eass

\brem \label{rem3}
According to Theorem 1.9 of \cite{huang3}, for $\forall E_i \in\mathbb R^{n_i\times q}, F_i\in\mathbb R^{p_i\times q}$ and $i\in\mathcal{\bar V}$, the following equations
\begin{equation}\label{eq28}
\begin{aligned}
	X_iS_0&=A_iX_i+B_iU_i+E_i\\
	0&=C_iX_i+D_iU_i+F_i
\end{aligned}
\end{equation}
 are solvable with unique solution $(X_i, U_i)$ if and only if Assumption~\ref{ass3} is satisfied. The equations in (\ref{eq28}) are usually called the regulator equations.
\erem
\brem\label{rem4}
Let $\Delta=\mbox{diag}\left\{a_{10},\cdots,a_{N0}\right\}$, then the Laplacian matrix $\mathcal{L}$ of  $\mathcal{G}$ can be written as
\begin{equation}\label{H}
	\mathcal{L}={\left [\begin{array}{c|c}
			\sum_{j=1}^Na_{0j}&-(a_{01},\cdots,a_{0N})\\[2mm]
			\hline
			-\Delta 1_N&H
		\end{array}
		\right ].
	}
\end{equation}
As shown  in \cite{huangjie1}, under Assumption~\ref{ass4}, $-H$ is Hurwitz and {\blue
\begin{equation}
	-(P_HH)\otimes I_q-(H\t P_H)\otimes I_q=-Q_H\otimes I_q.\label{PQ1}
\end{equation}}
holds with $P_H,Q_H\in \mathbb R^{N\times N}$ being positive definite matrices.
\erem



For $N=1$, we can ignore the subscript $i$ to simplify the presentation in this subsection.
Also, in this case, the leader state $\upsilon_0$ can be accessed by the only follower and the
observer \eqref{18} is not needed. As a result, we consider the state feedback controller
(\ref{eq11}) with $\upsilon_i$ replaced by $\upsilon_0$.
\bthm\label{the3}
{\blue Consider the systems (\ref{sys1}) and (\ref{sys2}) } with $N=1$ under Assumption \ref{ass3}. {\blue Let $\bar K$ be  any real matrix } and $\tilde K=U-\bar K X$ where $(X,U)$ satisfies (\ref{eq28}).	
Then, the PTOR problem is solvable  by employing  controller (\ref{eq11})  with $\upsilon_i$ replaced by $\upsilon_0$
if and only if
\begin{equation}\label{equa19}
	\max\left\{\Re(\lambda(BK))\right\}<-1.
\end{equation}
\ethm
\proofnow
(\textbf{\emph{Sufficiency}}) Define  $\bar{x}=x-X\upsilon_0$. Using  (\ref{sys1}), (\ref{sys2}) and (\ref{eq28}) gives
\begin{equation}\label{equa14}
	\dot{\bar{x}}= (A_{c} + \mu BK)\bar{x}.
\end{equation}
where $A_{c}=A+B\bar K$.
Let us introduce
 the time-varying state transformation
$\omega=\pi (t,m)\bar{x}$  where $\pi (t,m)=\exp\left(m\sint_{t_0}^t\mu(\tau )\mathrm d\tau \right)$
with a constant $m>0$. Note that $\omega$ is differentiable with respect to $t$.   Then, according to \eqref{equa14}, the $\omega$-dynamics and output  $e$ can be expressed as
\begin{equation}\label{equa6}
\begin{aligned}
	\dot{\omega}&=(A_{c} + \mu A_{k})\omega \\
	e&=(C_c +\mu DK )\bar{x}
\end{aligned}
\end{equation}
where $A_{k}=mI+BK$ and $C_{c}=C+D\bar K$. Solving (\ref{equa6}) yields
\begin{equation}\label{equa7}
	\omega(t)
	=\Phi_1(t)\Phi_2(t)\omega(t_0)
\end{equation}
where
\begin{gather*}
	\Phi_1(t)=\exp\left(A_{c}(t-t_0)\right)\\ \Phi_2(t)=\exp\left(A_{k}\sint_{t_0}^t\mu(\tau)\mathrm d\tau\right).
\end{gather*}
The singularity of the solution caused by the piecewise continuous function $\mu(t)$ can be addressed by the generalized Filippov solution  proposed in \cite{SEEBER91}.


Suppose that the Jordan canonical form of the matrix $A_{k}$ is denoted as  $J$ that
is composed of $r$ Jordan blocks, each of which has order $a_j$, i.e., $\sum_{j=1}^ra_j=n$.
In particular,  a nonsingular matrix $M$ can be found such that $A_{k}=MJM^{-1}$ and {\blue $J=\mbox{diag}\left\{J_{1}(\delta_1),\cdots,J_{r}(\delta_r)\right\}$, where $J_{j}(\delta_j)\in\mathbb R^{a_j\times a_j}$ is the $j$-th Jordan block with the eigenvalue $\delta_j$.}
The eigenvalues $\delta_j, j=1,\cdots,r$, are not necessarily distinct. Then
\begin{equation}\label{equa8}
\begin{aligned}
	\Phi_2(t)&=\sum_{j=0}^{\infty}\frac{1}{j!}\left(MJM^{-1}\sint_{t_0}^t\mu(\tau)\mathrm d\tau\right)^j\\
	&=M\left(\sum_{j=0}^\infty\frac{1}{j!}\left(J\sint_{t_0}^t\mu(\tau)\mathrm d\tau\right)^j\right)M^{-1}\\
	&=M {\blue\mathcal  J(t,\delta) }M^{-1}
\end{aligned}
\end{equation}
{\blue where $\delta = [\delta_1,\cdots,\delta_r]\t$. }
Since $J$ is in Jordan canonical form, the $j$-th diagonal block $ \mathcal J_{j}(t,\delta_j)$ of $\mathcal J(t,\delta)$ can be calculated as {\blue
\begin{align}
&\mathcal J_{j}(t,\delta_j)\notag\\
&=I_{a_j}+J_{j}(\delta_j)\sint_{t_0}^t\mu(\tau)\mathrm d\tau+\frac{1}{2}J_{j}^2(\delta_j)\left(\sint_{t_0}^t\mu(\tau)\mathrm d\tau\right)^2\notag\\
&\quad +\cdots+\frac{1}{l!}J_{j}^l(\delta_j)\left(\sint_{t_0}^t\mu(\tau)\mathrm d\tau\right)^l+\cdots\notag\\
&=\pi(t,\Re (\delta_j))\zeta_j(t,\delta_j)\mathcal{Z}_j(t,a_j)\label{equa9}
\end{align}
where we used
$
\sum_{l=0}^\infty \frac{1}{l!}\left(\delta_j\sint_{t_0}^t\mu(\tau)\mathrm d\tau\right)^l=\exp\left(\delta_j \sint_{t_0}^t\mu(\tau)\mathrm d\tau \right)
$
and
\begin{gather*}
\pi (t, \Re (\delta_j)) = \exp \left(\Re (\delta_j) \sint_{t_0}^t \mu(\tau) \mathrm d\tau\right),\\
\zeta_j(t,\delta_j)=\pi (t,i\Im (\delta_j))
	=\cos\left(\Im (\delta_j)\sint_{t_0}^t\mu(\tau)\mathrm d\tau)\right)\\
	+i\sin\left(\Im (\delta_j)\sint_{t_0}^t\mu(\tau)\mathrm d\tau\right),\\
	\mathcal{Z}_j(t,a_j)=
	\left [\begin{array}{cccc}
		1&\sint_{t_0}^t\mu(\tau)\mathrm d\tau &\cdots&\frac{\left(\sint_{t_0}^t\mu(\tau)\mathrm d\tau\right)^{a_j-1} }{(a_j-1)!}\\
		0&1&\cdots&\frac{\left(\sint_{t_0}^t\mu(\tau)\mathrm d\tau\right)^{a_j-2}  }{(a_j-2)!}\\
		\vdots&\vdots&\ddots&\vdots\\
		0&0&\cdots&1
	\end{array}
	\right ].
\end{gather*}
Therefore, $\mathcal J(t,\delta)$ can be expressed as
\begin{equation*}
\mathcal J(t,\delta)=\mbox{diag}\left\{\mathcal{J}_{1}(t,\delta_1),\cdots,\mathcal{J}_{r}(t,\delta_r)\right\}
\end{equation*}
in which each diagonal block has the form of \eqref{equa9}. }

Condition \eqref{equa19} implies the existence of positive constants $c_j$, $j=1,\cdots,r$, satisfying $\Re(\delta_j)=m-1-c_j$.
According to (\ref{equa6}), one has
\begin{align}\label{equa12}
	\| e(t)\| &\leq  (\|C_{c}\| + \mu\| DK\|)\|\bar{x}(t)\|\notag\\
	&\leq  (T\| C_{c}\|  + \| DK\|)\|\Phi_1(t)\|\| M\|\| M^{-1}\| \|\omega(t_0)\|  \notag\\
	&\quad   \times\| \mu(t)\pi(t,-m)\mathcal J(t,\delta)\|
\end{align}
where (\ref{equa7}), (\ref{equa8}) and $1/\mu(t)\leq T$ for $t\geq t_0$ are used in the calculation.
Note that
\begin{align}\label{equa13}
&\| \mu(t)\pi(t,-m)\mathcal J(t,\delta)\|\notag\\
&=\|\mbox{diag}\{\mu(t)\pi(t,-1-c_1)\zeta_1 (t,\delta_1)\mathcal{Z}_1(t,a_1),\notag\\
&\qquad\qquad \cdots,
	\mu(t)\pi(t,-1-c_r)\zeta_r(t,\delta_r)\mathcal{Z}_r(t,a_r)\}\|
\end{align}
{\blue  According to \eqref{equa9}, the elements  of $\mu(t)\pi(t,-1-c_j)\mathcal{Z}_j(t,a_j)$ equal to zero, or have the form $\frac{1}{\bar a_j!}\mu(t)\pi(t,-1-c_j)\left(\sint_{t_0}^t\mu(\tau)\mathrm d\tau \right)^{\bar{a}_j}$ for $\bar a_j = 0,\cdots,a_j-1$.}   Note that
\begin{align}
&\frac{1}{\bar a_j!}\mu(t)\pi(t,-1-c_j)\left(\sint_{t_0}^t\mu(\tau)\mathrm d\tau \right)^{\bar{a}_j}\notag\\
&\leq \Xi(t)\label{Xi}:=\frac{1}{\bar a_j!}\frac{1}{T}\pi(t,-c_j)\left(\sint_{t_0}^t\mu(\tau)\mathrm d\tau \right)^{\bar{a}_j}
\end{align}
where we used $\mu(t)\leq T^{-1}\pi(t,1)$ for $t\geq t_0$.
Expanding $\pi(t,-c_j)$ by Taylor series obtains
\begin{equation}\label{equa26}
\begin{aligned}
\Xi(t)&=\frac{1}{\bar a_j!}\frac{1}{T}\frac{\left(\sint_{t_0}^t\mu(\tau)\mathrm d\tau \right)^{\bar{a}_j}}{\sum_{l=0}^\infty\frac{1}{l!}(c_j)^l\left(\sint_{t_0}^t\mu(\tau)\mathrm d\tau \right)^{l}}\\
&=\frac{1}{\bar{a}_j!}\frac{1}{T}\left[\sum_{l=0}^\infty\frac{1}{l!}(c_j)^l \left(\sint_{t_0}^t\mu(\tau)\mathrm d\tau \right)^{l-\bar a_j}\right]^{-1}.
\end{aligned}
\end{equation}
Therefore,
for $t_0\leq t<T+t_0$, $\Xi(t)<\infty$  and $\lim_{t\to T+t_0}\Xi(t)=0$ due to {\blue
\begin{equation*}
\lim_{t\to (T+t_0)}\sint_{t_0}^t\mu(\tau)\mathrm d\tau=\ln\left(\textstyle{\frac{T}{T+t_0-t}}\right){\Big |}_{t_0}^{T+t_0}=\infty.
\end{equation*}}
For $t\geq T+t_0$, by \eqref{equa26}, we have
\begin{align}
\Xi(t)&=\frac{1}{\bar{a}_j!}\frac{1}{T}\left[\sum_{l=0}^\infty\frac{1}{l!}(c_j)^l \left(\infty +a(t-T-t_0) \right)^{l-\bar a_j}\right]^{-1}\notag\\
&=0 \notag
\end{align}
{\blue  where $a$ is introduced in \eqref{mu}. }

Therefore, by \eqref{equa13}, \eqref{Xi},  and the properties of $\Xi(t)$,
$\mu(t)\pi(t,-m)\mathcal J(t,\delta )$ is continuous for all $t\geq 0$, and converges to zero as $t\to T+t_0$, remains  zero afterwards.
Also note that $\Phi_1(t)<\infty$  for $t_0\leq t<T+t_0$. Therefore, by (\ref{equa12}), {\blue $\lim_{t \to T+t_0}e(t) =0$} and $e(t)=0$ for $t\geq T+t_0$.

(\textbf{\emph{Necessity}})
Suppose that the Jordan canonical form of the matrix $BK$ is denoted as $J'$ that is
composed of $h$ Jordan blocks, each of which has order $b_j$  for $j=1,\cdots, h$.
Note that $\sum_{j=1}^h b_j=n$. Then,  a nonsingular matrix $M'$ can be found such that the   solution of (\ref{equa14}) is
\begin{equation*}\label{equa15}
	\bar{x}(t)=\Phi_1(t)M' \mathcal {J}'(t,\iota )(M')^{-1}\bar{x}(t_0)
\end{equation*}
where   $\mathcal {J}'(t,\iota )=\mbox{diag}\left\{\mathcal {J}_{1}'(t,\iota_1),\cdots,\mathcal{J}_{h}'(t,\iota_h)\right\}$ with $ \iota = [\iota_1,\cdots,\iota_h]\t$ being the $h$-dimensional vector that contains all the eigenvalues of $BK$.
The function $\mathcal{J}_{j}'(t,\iota_j)$ has the similar form (\ref{equa9}), i.e., for $j=1,\cdots,h$,
\begin{equation}\label{eq95}
	\mathcal{J}_{j}'(t,\iota_j)=\pi(t,\Re (\iota_j))\zeta_j(t,\iota_j)\mathcal{Z}_j(t,b_j).
\end{equation}
Then, $e(t)$ can be expressed as
\begin{equation*}\label{eq98}
	e(t)=( C_c/\mu +DK)\Phi_1(t)M' \mu(t)\mathcal {J}'(t,\iota )M'^{-1}\bar{x}(t_0).
\end{equation*}
Since $\lim_{t\rightarrow t_0+T} e(t)=0$ and $e(t)=0,\, t\geq T+t_0$,  for  $\forall \bar x (t_0)\in\mathbb{R}^n$,  and
the matrices $\Phi_1(t)$ and $M'$ are non-singular, one must have
$\lim_{t\rightarrow t_0+T} \mu(t)\mathcal {J}'(t,\iota )=0$, and $ \mu(t)\mathcal {J}'(t,\iota )=0$ for all $t\geq T+t_0$.   By (\ref{eq95}),
\begin{equation}
	\begin{aligned}
	\mu(t)\bar{J}'(t,\iota)=&\mbox{diag}\left\{\mu(t)\pi(t,\Re (\iota_1))\zeta_1(t,\iota_1)\mathcal{Z}_1(t,b_1),\right.\\
	&\left.\qquad\cdots,
	\mu(t)\pi(t,\Re (\iota_h))\zeta_h(t,\iota_h)\mathcal{Z}_h(t,b_h)\right\}.
	\end{aligned}\notag
\end{equation}
As discussed in (\ref{equa13}) and (\ref{equa26}), since  $\lim_{t\to T+t_0}\mu(t)\mathcal {J}'(t,\iota )=0$  and $\mu(t)\mathcal {J}'(t,\iota )=0$ for $t\geq T+t_0$, one has
$-\Re(\iota_j)-1>0$, that is, $\Re(\iota_j)<-1$ for $j=1,\cdots,h$, which is equivalent to \eqref{equa19}.
\eproof

\medskip

The solvability of PTOR with the measurement output feedback control law can be similarly established, and its proof is omitted herein.
\bcorollary\label{coro:1}
Consider the systems (\ref{sys1}) and (\ref{sys2}) with $N=1$ under Assumption \ref{ass3}.
{\blue Let $\bar K$ and $L_1$ be  any real matrices.} Define $\tilde K=U-\bar K X$, where $(X,U)$ is the solution of (\ref{eq28}).	 Then  the PTOR problem is solvable  by employing controller in  \eqref{hatx}-\eqref{40} with $\upsilon_i$ replaced by $\upsilon_0$ if and only if both (\ref{equa19}) and
\begin{equation}\label{equa29}
	\min \left\{\Re (\tilde L C^\m)\right\}>1
\end{equation}
hold. 	\ecorollary

Note that  one of the solvability conditions of the output regulation problem with a state feedback control law is stabilizability of the pair $(A,B)$. Theorem \ref{the3} shows that  PTOR requires a  stronger solvability condition (\ref{equa19}).
The condition that there exists $K$ such that (\ref{equa19}) holds is equivalent to $\rank(B)=n$. If $\rank(B)=n$, it implies that   $(mI,B)$ is controllable and the eigenvalue of $BK$ can be freely allocated by $K$.
On the other hand, $\max \left\{\Re(\lambda(BK))\right\}<-1$   implies $\rank(BK)=n$. Note that $\rank(BK)\leq\min\left\{\rank(B),\rank(K)\right\}$, then we can conclude $\rank(B)=n$. Similarly,  the condition  that there exists an $\tilde L $ such that (\ref{equa29}) holds  is equivalent to $\rank(C^\m)=n$.
It explains that the following assumption is needed for PTOR and hence PTCOR to be studied in the subsequent sections.

\bass \label{ass2}
The matrices  $B_i$ and $C^\m_i $ satisfy $\rank(B_i)= \rank(C^\m_i  )=n_i$ for  $i\in\mathcal{\bar V}$.
\eass

{\blue
\brem
In \cite{chen2023prescribed97}, the sufficient condition of  solving the PTCOR   relies on   a set of LMIs. For instance, the state feedback approach needs to find     matrices $R_i>0$, $\bar K_i$, and $K_i$  such that  $R_i (A_i +B_i \bar K_i) + (A_i +B_i \bar K_i)\t R_i<0$ and $R_i B_i K_i + K_i\t B_i\t R_i<0$ hold. By Theorem 4.6 in \cite{2002Nonlinear99}, the second inequality implies that         $B_i K_i$ is Hurwitz, which is equivalent to that in   Theorem \ref{the3} and Corollary \ref{coro:1}. Theorem \ref{the3} and Corollary \ref{coro:1}     give     the condition which is  sufficient and necessary,  and it must be imposed for solving the PTCOR.
\erem
}

{\blue
\brem
Although Assumption \ref{ass2} is more stringent   than the conventional assumptions of $(A_i,B_i)$ being stabilizability and $(C_i^\m, A_i)$ being observability, we can find   practical systems  satisfying the condition.   For instance, the control problem of current-controlled voltage -source inverters (CCVSIs), as will be discussed in \cite{Caihe52}, can be reformulated into a COR problem of linear heterogeneous MAS satisfying Assumption \ref{ass2}.
\erem
}

\section{Prescribed-time Stabilization of Cascaded System}\label{CasSys}
In this section, we demonstrate the conversion of the PTCOR problem for the closed-loop MASs \eqref{sys1} into the prescribed-time stabilization problem of a cascaded system. Subsequently,  we propose a criterion of prescribed-time stabilization for the cascaded system.
\subsection{State Transformation and Error System}
For $i\in \bar{\mathcal V }$,  define
\begin{equation}\label{StaTrans}
	\begin{aligned}
		\tilde \upsilon_i=\upsilon_i-\upsilon_0 ,\quad \bar x_i=x_i-X_i \upsilon _0, \quad \tilde x_i=\hat x_i-x_i
	\end{aligned}
\end{equation}
as  the estimator error for the distributed observer, local state tracking error, and   estimate error for the local state observer, respectively.
Denote the lumped vector variables
\begin{equation}
\begin{gathered}
	\tilde{\upsilon}=\left[\tilde{\upsilon}_1\t,\cdots,\tilde{\upsilon}_N\t\right]\t,\quad \bar x=\left[\bar x_1\t,\cdots,\bar x_N\t\right]\t,\\
\tilde x=\left[\tilde x_i\t,\cdots,\tilde x_N\t\right]\t,\quad e=[e_1\t,\cdots,e_N\t]\t.
\end{gathered}
 \end{equation}
 By \eqref{18}, (\ref{H}) and  $(\Delta \otimes I_q)(1_N\otimes \upsilon_0)=(H\otimes I_q)(1_N\otimes\upsilon_0)$,
the ${\tilde{\upsilon}}$-dynamics  can be expressed as {\blue
\begin{equation}\label{19}
	\begin{aligned}
		\dot{\tilde{\upsilon}}& = \dot \upsilon -1_N \otimes \dot \upsilon_0 \\
& = -\psi \mu \left(\left[\begin{array}{cc}-\Delta 1_N & H\end{array}\right]\otimes I_q\right)\left[ \begin{array}{cc} \upsilon_0 \t &\upsilon\t\end{array}\right ]\t\\
&\quad + (I_N \otimes S_0) \upsilon- 1_N \otimes S_0\upsilon_0\\
&=-\psi\mu(H\otimes I_q)\upsilon+\psi\mu(\Delta\otimes I_q)(1_N\otimes \upsilon_0)\\
		&\quad +(I_N\otimes S_0)\tilde{\upsilon}\\
		&=-\psi\mu(H\otimes I_q)\upsilon+\psi\mu (H\otimes I_q)(1_N\otimes \upsilon_0) \\
		&\quad +(I_N\otimes S_0)\tilde{\upsilon}\\
		&=\left[I_N\otimes S_0-\psi\mu(H\otimes I_q)\right]\tilde{\upsilon}
	\end{aligned}
\end{equation}
where we used $1_N \otimes S_0\upsilon_0 = (I_N \otimes S_0)(1_N \otimes \upsilon_0)$ and $(\Delta I_N\otimes I_q)\upsilon_0 = (\Delta \otimes I_q) (1_N \otimes \upsilon_0)$.
}

For state feedback, the  $\bar x$-dynamics and regulated output $e$ are {\blue
\begin{align}
\dot {\bar x}&= \dot x -X (1_N \otimes \dot \upsilon_0)\notag\\
& = A x +Bu + E(1_N \otimes \upsilon_0) - X (1_N \otimes S_0\upsilon_0)\notag \\
& = Ax + B(\bar K x +\tilde K \upsilon +\mu K (x-X\upsilon))\notag\\
&\quad + E(1_N \otimes \upsilon_0) - X (1_N \otimes S_0\upsilon_0)\notag \\
& = (A_c+\mu BK)\bar x+(B\tilde K-\mu BKX)\tilde \upsilon\notag \\
&\quad + (AX+B\bar K  + B\tilde K +E -X (I_N\otimes S_0) )(1_N \otimes \upsilon_0)
\notag\\&=(A_c+\mu BK)\bar x+(B\tilde K-\mu BKX)\tilde \upsilon\label{domegai1}\\
e&= C x + Du + F(1_N \otimes \upsilon_0)\notag \\
& = Cx + D(\bar K x +\tilde K \upsilon + \mu K (x-X\upsilon))+F(1_N \otimes \upsilon_0) \notag\\
& = (C_c+\mu DK)\bar x+D(\tilde K-\mu KX)\tilde \upsilon  \notag \\
& \quad + (CX + D\bar K X + D\tilde K + F)(1_N\otimes \upsilon_0)\notag \\
&=(C_c+\mu DK)\bar x+D(\tilde K-\mu KX)\tilde \upsilon\label{eq:e_sta}
\end{align}
where $X=\mbox{diag}\{X_{1},\cdots,X_{N}\}$,
$A=\mbox{diag}\{A_{1},\cdots,A_{N}\}$,
$B=\mbox{diag}\{B_{1},\cdots,B_{N}\}$,
$u=[u_1\t,\cdots,u_N\t]\t$,
$E=\mbox{diag}\{E_{1},\cdots,E_{N}\}$,
$\bar K =\mbox{diag}\{\bar K_{1},\cdots,\bar K_{N}\}$,
$\tilde  K =\mbox{diag}\{\tilde K_{1},\cdots,\tilde  K_{N}\}$,
$ K =\mbox{diag}\{ K_{1},\cdots, K_{N}\}$,
$A_c=\mbox{diag}\{A_{c1},\cdots,A_{cN}\}$ with $A_{ci} = A_i + B_i \bar K_i$ for $i\in \bar {\mathcal V}$,
$C=\mbox{diag}\{C_{1},\cdots,C_{N}\}$,
$D=\mbox{diag}\{D_{1},\cdots,D_{N}\}$,
$F=\mbox{diag}\{B_{1},\cdots,F_{N}\}$, and
$C_c=\mbox{diag}\{C_{c1},\cdots,C_{cN}\}$ with $C_{ci} = C_i + D_i \bar K_i$.
}

For measurement output feedback,
taking time derivative of $\tilde x_i$ and using (\ref{sys1}) and (\ref{hatx}) obtain
\begin{gather}
	\dot{\tilde x}_i= \dot{\hat{x}}_i-\dot{x}_i
	=(A_i-L_{i}C^\m_i+\mu A_{Li})\tilde x_i\notag\\
	 +(E_i-L_{i}F^\m_i  -\mu \tilde L_{i}F^\m_i  )\tilde{\upsilon}_i\label{eq:dot_til_x}
\end{gather}
where $A_{Li}=-\tilde L_{i}C_i^\m$.
Then the  $\tilde x$-dynamics, $\bar x$-dynamics, and regulated output $e$ can be expressed as
\begin{align}
\dot{\tilde x}&=(A-LC^\m +\mu A_L)\tilde x\notag \\
&\quad+(E-LF^\m-\mu \tilde L F^\m)\tilde \upsilon\label{dot_til_x} \\
\dot{\bar x}&=(A_c+\mu A_K)\bar x+(B\bar K+\mu BK)\tilde x\notag\\
&\quad+(B\tilde K-\mu BKX)\tilde \upsilon\label{dot_bar_x}\\
e&=(C_c+\mu DK)\bar x+D(\tilde K -\mu KX)\tilde \upsilon \notag\\
&\quad+D(\bar K+\mu K)\tilde x.\label{e_2}
\end{align}
where  $L=\mbox{diag}\{L_{1},\cdots,L_{N}\}$,  $C^\m=\mbox{diag}\{C^\m_1,\cdots,C^\m_N\}$,  $A_L=\mbox{diag}\{A_{L1},\cdots,A_{LN}\}$,  $F^\m=\mbox{diag}\{F^\m_1,\cdots,F^\m_N\}$, $\tilde L=\mbox{diag}\{\tilde L_{1},\cdots,\tilde L_{N}\}$, and the other matrixes are same as that in \eqref{domegai1} and \eqref{eq:e_sta}.

By examining \eqref{domegai1}-\eqref{e_2},
the error systems can be succinctly expressed as a cascaded system
\begin{equation}\label{chisys}
	\dot \chi_1=f_1(t,\chi_1),\;\;
		\dot \chi_2=f_2(t,\chi_1,\chi_2),\;\; e=h(t,\chi_1,\chi_2)
\end{equation}
 where for state feedback
$
 \chi_1=\tilde \upsilon$, $\chi_2=\bar x,
$
 and for measurement output feedback
$
 \chi_1=\left[\tilde \upsilon\t,\tilde x\t\right]\t$, $\chi_2=\bar x.
$

\subsection{Cascaded System}
A criterion of  prescribed-time convergence for the   cascaded system in the form of (\ref{chisys}) is proposed, which  holds significant importance in analyzing the PTCOR implementation of
closed-loop MASs \eqref{sys1}.

	\begin{lemma}\label{lem:CasSysSta}
		Suppose the dynamics of $\chi_1$ and $\chi_2$ in (\ref{chisys}) admit the PTLF and PTISSLF  in Definition \ref{def:PTISSLF}, respectively, i.e., there exist  Lyapunov functions $V_1(\chi_1)$, $V_2(\chi_2)$ such that
\begin{gather}
\begin{gathered}
\underline {\alpha }_1\|\chi_1\|^2\leq V_1(\chi_1)\leq \bar \alpha_1 \|\chi_1\|^2\\
\dot V_1(\chi_1) \leq -\alpha_1 \mu V_1 (\chi_1)+\tilde \alpha_1 V_1(\chi_1)
\end{gathered}\label{eq:V_1}\\
\begin{gathered}
\underline {\alpha }_2\|\chi_2\|^2\leq V_2(\chi_2)\leq \bar \alpha_2 \|\chi_2\|^2\\
\dot V_2 (\chi_2)\leq -\alpha_2 \mu V_2 (\chi_2)+\tilde \alpha_2 V_2(\chi_2)+\sigma \mu^m\|\chi_1\|^n.
\end{gathered}\label{eq:V_2}
\end{gather}
		Suppose
\begin{equation}
\| e(t)\|\leq  \varepsilon_e \mu^p(t)\| \chi(t)\|\label{eq:e}
\end{equation}
holds for $t\geq t_0$ and some positive finite constant $ \varepsilon_e$ and $p$. For a given $\alpha^*$, if $\alpha_1$, $\alpha_2$ satisfy
\begin{equation}\label{eq:eta}
\begin{gathered}
\alpha_2\geq 2(p+\alpha^*)\\
 \alpha_1\geq \max\{2(\alpha_2+m)/n,2(p+\alpha^*)\}
\end{gathered}
\end{equation}
 then $\chi:=[\chi_1\t,\chi_2\t]\t$ and $e$ converge to zero within the prescribed time and remain as zero afterwards. In particular,  $\mathcal{K}$ functions $\gamma_\chi$, $\gamma_e$ and a constant $\tilde \alpha$  can be found to yield
\begin{align}
\|\chi(t)\|\leq& \gamma_\chi(\|\chi(t_0)\|)\kappa^{p+\alpha^*}(t-t_0) \exp(\tilde \alpha(t-t_0))\label{beta_chi}\\
\|e(t)\|\leq& \gamma_e(\|\chi(t_0)\|)\kappa^{\alpha^*}(t-t_0) \exp(\tilde \alpha(t-t_0)).\label{beta_e}
\end{align}
with
\begin{align}
	\kappa(t-t_0)&=\exp\left(-\sint_{t_0}^t\mu(\tau)\mathrm d\tau\right)\notag\\
	&=\left\{\begin{array}{c}\frac{T+t_0-t}{T},\\0,\end{array}\right.\begin{array}{c}t_0\leq t<T+t_0\\T+t_0\leq t\end{array}.\label{kappa}
\end{align}
	\end{lemma}

	\proofnow 	
	 Invoking comparison lemma for the second inequality of \eqref{eq:V_1} yields
	\begin{equation}
		V_1(\chi_1(t))\leq \kappa^{\alpha_1}(t-t_0)\exp (\tilde\alpha_1(t-t_0))V_1(\chi_1(t_0))\notag
	\end{equation}
	Then $\chi_1$ satisfies
	\begin{align}
		\|\chi_1(t)\|\notag&\leq\sqrt{\bar\alpha_1/\underline \alpha_1}\|\chi_1(t_0)\|\notag\\
&\quad \times\kappa^{\frac{\alpha_1}{2}}(t-t_0)\exp\left(\frac{\tilde\alpha_1}{2}(t-t_0)\right)
.\label{beta_1}
	\end{align}
Due to the property of $\kappa$ function in \eqref{kappa},
we note $\lim_{t\rightarrow T+t_0}\chi_1(t)=0$ and $\chi_1(t)=0$ for $t\geq T+t_0$.
Invoking comparison lemma for the second inequality in \eqref{eq:V_2} yields
	\begin{align}
		V_2(\chi_2(t))&\leq \kappa^{\alpha_2}(t-t_0)\exp (\tilde \alpha_2(t-t_0))V_2(\chi_2(t_0))\notag\\
		&\quad+\sint_{t_0}^t\exp\left(-\sint_{\tau}^t\alpha_2\mu(s)\mathrm ds+\tilde \alpha_2(t-\tau)\right)\notag\\
		&\quad \times \sigma \mu^m(\tau)\|\chi_1(\tau)\|^n\mathrm d\tau.
	\end{align}
For $\mu (t)$ in \eqref{mu}, one has
\begin{equation}
\mu(t)\leq T^{-1}\kappa^{-1}(t-t_0),\quad \forall t\geq t_0.\label{eq:mu1}
\end{equation}
By (\ref{eq:eta}), we have $\frac{\alpha_1n}{2}\geq \alpha_2+m$.
 Then
by (\ref{beta_1}),	the second term on the right-hand side of the inequality can be calculated as
	\begin{align}
		&\sint_{t_0}^t\exp\left(-\sint_{\tau}^t\alpha_2\mu(s)\mathrm ds+\tilde \alpha_2(t-\tau)\right)\sigma \mu^m(\tau)\|\chi_1(\tau)\|^n\mathrm d\tau\notag\\
&\leq d_1\kappa^{\alpha_2}(t-t_0)\exp (\tilde\alpha_2 (t-t_0))\sint_{t_0}^t \kappa^{-\alpha_2}(\tau-t_0) \notag\\
&\quad \times \mu^m(\tau)\kappa^{\frac{\alpha_1 n}{2}}(\tau-t_0)\exp\left(\frac{\tilde\alpha_1 n}{2}(\tau-t_0)\right)\mathrm d\tau\notag\\
&\leq d_1T^{-m}\kappa^{\alpha_2}(t-t_0)\exp\left (\left(\tilde\alpha_2+\frac{\tilde\alpha_1 n}{2}\right) (t-t_0)\right)\notag\\
&\quad \times \sint_{t_0}^t\kappa^{\frac{\alpha_1 n}{2}-m-\alpha_2}(\tau-t_0)\mathrm d\tau\notag\\
&\leq d_1T^{-m}\kappa^{\alpha_2}(t-t_0)\exp\left (\tilde \alpha_2' (t-t_0)\right)
		\label{SecTerm2}
	\end{align}
where $d_1=\sigma\|\chi_1(t_0)\|^n\left(\frac{\bar \alpha_1}{\underline \alpha_1}\right)^{\frac{n}{2}}$, $\tilde\alpha'_2=\tilde\alpha_2+\frac{\tilde\alpha_1 n}{2}+1$, and we used the facts
\begin{gather}
\exp\left(-\sint_{\tau}^t\alpha_2\mu(s)\mathrm ds\right)=\frac{\kappa^{\alpha_2}(t-t_0)}{\kappa^{\alpha_2}(\tau-t_0)}\notag\\
\sint_{t_0}^t\kappa^{\frac{\alpha_1 n}{2}-m-\alpha_2}(\tau-t_0)\mathrm d\tau\leq \sint_{t_0}^t 1\mathrm d\tau\leq \exp(t-t_0).\notag
\end{gather}
	Therefore, $V_2(\chi_2 (t))$ satisfies
	\begin{align}
		V_2(\chi_2(t))&\leq \kappa^{\alpha_2}(t-t_0)\exp(\tilde\alpha_2'(t-t_0))\notag\\
&\quad \times(V_2(\chi_2(t_0))+d_2\|\chi_1(t_0)\|^n)
	\end{align}
where $d_2=T^{-m}\sigma\left(\frac{\bar \alpha_1}{\underline \alpha_1}\right)^{\frac{n}{2}}$.
	As a result, the bound of  $\chi_2$ satisfies
	\begin{align}
		\|\chi_2(t)\|&\leq \sqrt{(\bar\alpha_2\|\chi_2(t_0)+d_2\|\chi_1(t_0)\|^n)/\underline \alpha_2}\notag\\
&\quad \times \kappa^{\frac{\alpha_2}{2}}(t-t_0)\exp\left(\frac{\tilde\alpha_2'}{2}(t-t_0)\right).\label{eq:chi_2}
	\end{align}
Therefore,   $\lim_{t\rightarrow T+t_0}\chi_2(t)=0$ and $\chi_2(t)=0$ for $t\geq T+t_0$.
By (\ref{eq:eta}), one has $\min\{\alpha_1,\alpha_2\}\geq 2(p+\alpha^*)$.
According to (\ref{beta_1}) and (\ref{eq:chi_2}),  one has  (\ref{beta_chi}) is proved for $\tilde  \alpha=\frac{1}{2}\max\{\alpha_2,\tilde \alpha_2'\}$ and any $\gamma_\chi\in\mathcal K$ satisfying
\begin{align}
\gamma_\chi(\|\chi(t_0)\|)&\geq \sqrt{\bar\alpha_1/\underline \alpha_1}\|\chi_1(t_0)\|\notag\\
&\quad +\sqrt{(\bar\alpha_2\|\chi_2(t_0)+d_2\|\chi_1(t_0)\|^n)/\underline \alpha_2}.
\end{align}

By (\ref{eq:e}), (\ref{beta_chi}) and (\ref{eq:mu1}), $e(t)$ satisfies
\begin{equation*}
\|e(t)\|
\leq \varepsilon_e T^{-p}\gamma_\chi(\|\chi(t_0)\|)\kappa^{\alpha^*}(t-t_0) \exp(\tilde \alpha(t-t_0)).
\end{equation*}
Then, \eqref{beta_e} is proved for $\gamma_e(\|\chi(t_0)\|)=\varepsilon_eT^{-p}\gamma_\chi(\|\chi(t_0)\|)$.
	\eproof

{\blue
\brem
Since the prescribed-time convergent rate of $\chi_1$  affects the prescribed-time stability of $\chi_2$-dynamics, the criterion  \eqref{eq:eta} implies that
the gain design of the $\chi_1$-dynamics must consider  the gain from  $\chi_2$-dynamics in order to achieve the prescribed-time stabilization of the cascaded system, which is different from  the
asymptotic stabilization \cite{panteley2001growth94, nguyen2024formation96} or finite-time stabilization \cite{ding2012nonsmooth93,wang2024fixed95} of a cascaded system.
\erem
}

\section{Stability Analysis}\label{sec:StaAna}
In this section, we establish that with suitable parameter choices, the whole closed-loop MASs satisfy the conditions of Lemma \ref{lem:CasSysSta}, thereby achieving PTCOR using both state feedback and measurement output feedback methods.
\subsection{Distributed Observer}
\blem \label{lem:do}
 Consider the distributed observers (\ref{18}) and exosystem (\ref{sys2}) under Assumption  \ref{ass4}.
If $\psi$  is sufficiently  large such that
\begin{equation*}
\psi\rho_H > 1
\end{equation*}  for
	\begin{equation} \label{rhoH}
		\rho_H=\frac{1}{2}\lambda_{\min}(Q_H)\lambda_{\max}^{-1}(P_H)
	\end{equation}
	where $P_H$ and $Q_H$ are defined in \eqref{PQ1}.    Then the $\tilde \upsilon$-dynamics admits a
	PTLF  satisfying \eqref{eq:V_1} in Lemma \ref{lem:CasSysSta} with
	\begin{equation}\label{con1}
		\begin{aligned}
			&\underline\alpha_1=\lambda_{\min}(P_H),\, \bar\alpha_1=\lambda_{\max}(P_H)\\
			&\alpha_1=2\psi\rho_H ,\, \tilde\alpha_1=\varpi
	\end{aligned}\end{equation}
where
$
	\varpi=2\|P_H\|\|S_0\|\lambda_{\min}^{-1}(P_H)
$.
	Moreover,
	$\tilde \upsilon$ is bounded for $t\geq t_0$ and converges to zero at $T+t_0$, remains as zero afterwards. Additionally,  $ \phi_1$ in \eqref{eq:phi_signal}  is bounded.
\elem
\proofnow
Since $H$ is Hurwitz, by Remark \ref{rem4}, define
\begin{equation}\label{V_upsilon}
V(\tilde \upsilon)= \tilde \upsilon\t( P_H\otimes I_q)\tilde \upsilon.
\end{equation}
Then its time derivative along trajectory of (\ref{19}) is
\begin{align}\label{dot_V_upsilon}
\dot V(\tilde \upsilon)=&-\psi \mu \tilde \upsilon \t(Q_H\otimes I_q)\tilde \upsilon+2\tilde \upsilon\t(P_H\otimes S_0)\tilde \upsilon \notag\\
\leq & -2\psi \rho_H\mu V(\tilde \upsilon)+\varpi V(\tilde \upsilon)
\end{align}
where we used $(P_H\otimes I_q)(I_N\otimes S_0)=P_H\otimes S_0$. Then (\ref{V_upsilon}) and (\ref{dot_V_upsilon}) satisfy \eqref{eq:V_1}.
Following the methodology used in proving Lemma \ref{lem:CasSysSta}, invoking the comparison lemma for (\ref{dot_V_upsilon}) obtains
\begin{align}
\|\tilde \upsilon(t)\|&\leq \sqrt{\lambda_{\max}(P_H)/\lambda_{\min}(P_H)}\|\tilde \upsilon(t_0)\|\notag\\
&\quad \times \kappa^{\psi \rho_H}(t-t_0)\exp \left(\frac{\varpi}{2}(t-t_0)\right).\label{til_upsilon}
\end{align}
What is left is to prove that $\phi_1(t)$ in \eqref{eq:phi_signal} is bounded $t\geq t_0$. Noting  $\upsilon_j-\upsilon_i =\tilde \upsilon_j-\tilde \upsilon_i$,
it is sufficient to demonstrate  that $\mu(t)\tilde\upsilon_i(t)<\infty$ for $t\geq t_0$.
Indeed, by \eqref{til_upsilon}, one has
\begin{equation}\label{mu_til_upsilon}
\| \mu(t)\tilde\upsilon_i(t) \| \leq \varepsilon_\upsilon\kappa^{\psi \rho_H-1}(t-t_0)\exp \left(\frac{\varpi}{2}(t-t_0)\right)
\end{equation}
for some constant $\varepsilon_\upsilon$. We note the term in the right-hand of \eqref{mu_til_upsilon}
 is bounded for $\psi\rho_H-1 >0$.
\eproof

\subsection{PTCOR with State Feedback}

In this subsection, we delve into the PTCOR problem utilizing the distributed observer (\ref{18}) and the state feedback controller (\ref{eq11}). The main result is articulated in the following theorem, which includes the explicit construction of design parameters.



\bthm\label{the1} Consider the closed-loop system composed of  the MASs (\ref{sys1}), the exosystem (\ref{sys2}), the observer (\ref{18}), and the
state feedback controller (\ref{eq11}) under Assumptions~\ref{ass3}, \ref{ass4}, and \ref{ass2}. For $i\in\mathcal{\bar V}$, suppose the parameters are selected as follows,
\begin{itemize}
	\item {\blue $\bar K_i$ is any real matrix;}
	
	\item $\tilde K_i=U_i-\bar K_iX_i$ where $(X_i,U_i)$ satisfies (\ref{eq28});
	\item $K_i$ is such that $B_iK_i$ is Hurwitz and
	\begin{equation}  \label{eq:theta_i}
		\theta_i=\frac{1}{2}\lambda_{\min}(Q_{Ki})\lambda_{\max}^{-1}(P_{Ki}) >1
	\end{equation}
	where $P_{Ki}$ and $Q_{Ki}$ are positive definite matrices satisfying $P_{Ki}B_iK_i+(B_iK_i)\t P_{Ki}=-Q_{Ki}$; and
	\item $\psi$  is sufficiently  large such that
\begin{equation}\label{eq:psi_rho}
\psi\rho_H \geq \theta_i+1
\end{equation} where $\rho_H$ is given in (\ref{rhoH}).
\end{itemize}
Then,
the PTCOR problem is solved in the sense that
the regulated output $e_i$ achieves prescribed-time convergence
towards zero at $T+t_0$ and remains as zero aftherwards.
Moreover, the internal signals in the closed-loop system and the $\mu(t)$-dependent terms $\phi_1(t)$ and $\phi_2(t)$ in (\ref{eq:phi_signal})
are bounded for all $t\geq t_0$. \ethm

\proofnow
First, we can always find the matrices $\bar K_i$, $\tilde K_i$ and $K_i$ under
Assumptions~\ref{ass3} and \ref{ass2}.
Moreover, the matrix $K_i$  such that $\theta_i>1$ in (\ref{eq:theta_i}) can always be found under Assumption \ref{ass2}.
In particular, $K_i$ can be chosen as $K_i=-B_i^{-1}\bar{m}I$ with $\bar m>1$ being a constant. Letting $P_{Ki}=I$ implies  $Q_{Ki}=2\bar{m}I$. Then, we have $\theta_i=\bar{m}>1$.

We note that the closed-loop system is compactly expressed  in  \eqref{19}, \eqref{domegai1}, and \eqref{eq:e_sta}.
 For  $\bar x$-dynamics  \eqref{domegai1}, define
\begin{equation}\label{W_i}
W(\bar x)={\bar x}\t P_{K}\bar x
\end{equation}
where $P_K=\mbox{diag}\{P_{K1},\cdots,P_{KN}\}$ is positive definite. Then, $\dot W(\bar x)$ satisfies
\begin{align}\label{dot_W_i}
\dot W(\bar x) =& -\mu {\bar x}\t Q_{K}\bar x+2{\bar x}\t P_{K}A_{c}\bar x\notag\\
&+2{\bar x}\t P_{K}(B_i\tilde K-\mu BKX)\tilde{\upsilon}\notag\\
\leq &-2\mu \theta W(\bar x)+\varpi_{1} W(\bar x)+ \varpi _{2}\mu^2\|\tilde \upsilon\|^2
\end{align}
where $Q_K=\mbox{diag}\{Q_{K1},\cdots,Q_{KN}\}$ and we used Young's inequality, and
\begin{align}
&\theta =\min \{\theta_1,\cdots,\theta_N\}\notag\\
&\varpi_{1}=(2\|P_{K}\|\|A_{c}\|+2)\lambda_{\min}^{-1}(P_{k})\notag\\
&\varpi_{2}=\|P_{K}\|^2 \|B\|^2(\|\tilde K\|^2T^2+\|K\|^2\|X\|^2).\notag
\end{align}
Therefore, $\bar x$-dynamics  \eqref{domegai1} admits a PTISSLF   satisfying (\ref{eq:V_2}) in  Lemma \ref{lem:CasSysSta} with
\begin{equation}\label{eta_3}
	\begin{aligned}
		&\underline\alpha_2=\lambda_{\min}(P_{K}),\, \bar \alpha_2=\lambda_{\max}(P_{K})\\
		&\alpha_2=2\theta ,\,\tilde\alpha_2=\varpi_1,\, \sigma=\varpi_2,\,m=2,\,n=2.
	\end{aligned}
\end{equation}
The  regulated output error $e$ in \eqref{eq:e_sta} satisfies
\begin{align}
\|e(t)\|&\leq (T\|C_{c}\|+\|DK\|)\mu (t)\|\bar x(t)\|\notag\\
&\quad +(T\|D\tilde K\|+\|DKX\|)\mu(t)\|\tilde\upsilon (t)\|\label{eq:e_1}
\end{align}
which coincides with (\ref{eq:e}) in Lemma \ref{lem:CasSysSta} with
$\varepsilon_e=\max\{T\|C_{c}\|+\|DK\|,T\|D\tilde K\|+\|DKX\|\}$ and $p=1$.

By (\ref{con1}) and (\ref{eta_3}), we can prove (\ref{eq:eta}) is satisfied with
\begin{equation}
\alpha^*=\theta-1>0.
\end{equation}
As a result, all conditions of Lemma \ref{lem:CasSysSta} are satisfied. {\blue Let $\chi(t_0) = [\tilde\upsilon(t_0)\t, \bar x(t_0)\t]\t$,} then by \eqref{beta_chi}, \eqref{beta_e}, $\bar x$ and $e$ satisfy
\begin{equation}\label{e_x}
\begin{aligned}
\|\bar x(t)\|&\leq  \gamma_{\chi}(\|\chi(t_0)\|)\kappa^{\theta}(t-t_0) \exp(\tilde \alpha(t-t_0))\\
\|e(t)\|&\leq \gamma_{e }(\|\chi(t_0)\|)\kappa^{\alpha^*}(t-t_0) \exp(\tilde\alpha(t-t_0))
\end{aligned}
\end{equation}
for some $\gamma_{\chi }$, $\gamma_{e}\in \mathcal K$ and some positive finite constant $\tilde \alpha$.
The PTCOR problem is thus solved by noting $\|e_i(t)\|\leq \|e(t)\|$. {\blue Define
\begin{equation}
\tilde u_i = u_i - U_i \upsilon_0 \label{eq:tilde_u}
\end{equation}
as the tracking error for local controller $u_i$, and $\tilde u = [\tilde u_1 \t,\cdots,\tilde u_N\t]\t$ as the lumped vector. According to \eqref{eq11} and \eqref{StaTrans}, $\tilde u$ can be expressed as
\begin{align}
\tilde u = \bar K (\bar x -X\tilde \upsilon) +\mu (\bar x -X\tilde \upsilon ) +U\tilde \upsilon. \notag
\end{align}
Then, by \eqref{e_x}, the  bound of $\tilde u$ is
\begin{equation}
\|\tilde u(t)\|\leq \gamma_{\tilde u}(\|\chi(t_0)\|)\kappa^{\alpha^*}(t-t_0) \exp(\tilde\alpha(t-t_0))\label{eq:bound-tilde_u}
\end{equation}
where $\gamma_{\tilde u} \in \mathcal K$. } Establishing the boundedness of $v_0$ for all $t\geq t_0$ is straightforward since it is generated by a neutrally stable linear system. {\blue Consequently, the states $v_i$ and $x_i$, and controller $u_i$ of the closed-loop system remain bounded, as indicated by \eqref{til_upsilon},  \eqref{e_x} and \eqref{eq:bound-tilde_u}. }

According to Lemma \ref{lem:do},   $\phi_1(t)<\infty$ for $t\geq t_0$.
Noting $\mu(t)\|\bar x_i(t)\|\leq \mu(t)\|\bar x(t)\|$ and
$x_i-X_{i}\upsilon_i=\bar x_i - X_i \tilde \upsilon_i$, to prove  $\phi_2(t)<\infty$ for $t\geq t_0$, it is sufficient to demonstrate that   $\mu(t)\bar x(t)$ is bounded.
Indeed, by \eqref{e_x}, one has
\begin{eqnarray}\label{mutermupsilon}
\begin{aligned}
	\| \mu(t)\bar x(t) \| \leq   \varepsilon_{\bar x}\kappa^{\theta-1}(t-t_0)\exp \left(\tilde \alpha(t-t_0)\right)
\end{aligned}
\end{eqnarray}
for some constant  $\varepsilon_{\bar x}$. By \eqref{mutermupsilon}, $\phi_2(t)$
 is bounded for  $\theta-1>0$.
	\eproof
	
	\subsection{PTCOR with Measurement Output Feedback}\label{sec4}
	We first show the prescribed-time convergence of the local estimation error.
	\blem \label{lem5}
	Consider the  closed-loop system composed of  the MASs (\ref{sys1}), the exosystem (\ref{sys2}), the distributed observer (\ref{18}), and the
	local state observer (\ref{hatx}) under Assumption~\ref{ass2}. Suppose the parameters are selected as follows,
	for $i\in\mathcal{\bar V}$,
	\begin{itemize}

		\item   {\blue $L_{i}$ is any real matrix;}  and

		\item   $\tilde L_{i}$ is such that  $A_{Li}=-\tilde L_{i}C^\m_i  $ is Hurwitz and
		\begin{equation} \label{eq:vartheta_i}
			\vartheta_i=\frac{1}{2}\lambda_{\min}(Q_{Li})\lambda_{\max}^{-1}(P_{Li}) >1
		\end{equation}
		where $P_{Li}$ and $Q_{Li}$  are positive definite matrices satisfying
		$P_{Li}A_{Li}+A_{Li}\t P_{Li}=-Q_{Li}$.
\item  $\psi$ is sufficiently large such that
\begin{equation}\label{eq:psi_rho_1}
\psi\rho_H\geq \vartheta_i+1
\end{equation}
where $\rho_H$ is given in (\ref{rhoH}).
	\end{itemize}
Then, there exists a PTISSLF   with $\tilde \upsilon$ as the input for  $\tilde x$-dynamics in \eqref{dot_til_x},
	and  the local state estimation error
	$\tilde{x}_i$ for $i\in\bar {\mathcal V}$
	converges
	to zero at $T+t_0$ and remains as zero afterwards. Moreover,  the $\mu(t)$-dependent terms $\phi_1(t)$ and $\phi_3(t)$ in (\ref{eq:phi_signal})
	are bounded for  $t\geq t_0$.
	\elem
	
	\proofnow
	First, we can always find the matrix $\tilde L_{i}$ under Assumption \ref{ass2}.
	Indeed, $\tilde L_{i}$ can be chosen as $\tilde L_{i}=\bar{m}I(C^\m_i)^{-1}$ with $\bar{m}>1$ being a constant. Letting $P_{Li}=I$ obtains $Q_{Li}=2\bar{m}I$. Then, we have $\vartheta_i=\bar{m}>1$.
 For 	$\tilde x$-dynamics in \eqref{dot_til_x}, define
\begin{equation*}
\tilde V(\tilde x)=\tilde x\t P_{L}\tilde x
\end{equation*}
 with $P_L=\mbox{diag}\{P_{L1},\cdots,P_{LN}\}$ is positive definite.
Then, $\dot {\tilde V} (\tilde x)$ satisfies
\begin{equation}\label{dot_V_i}
\dot {\tilde V}(\tilde x)\leq -2\vartheta \mu \tilde V(\tilde x)+\varpi_{3}\tilde V(\tilde x)+\varpi_{4}\mu^2\|\tilde \upsilon\|^2
\end{equation}
where
\begin{equation*}
\begin{aligned}
&\vartheta=\min\{\vartheta_1,\cdots,\vartheta_N\}\\
&\varpi_{3}=(2\|P_{L}\|\|A-L_{1}C^\m\|+2)\lambda_{\min}^{-1}(P_{L})\\
&\varpi_{4}=\|P_{L}\|^2(\|E-L_{1}F^\m\|^2T^2+\|\tilde L F^\m\|^2).
\end{aligned}
\end{equation*}
Therefore,  $V(\tilde x)$ is a PTISSLF $\tilde x$-dynamics in \eqref{dot_til_x}. Invoking Lemma \ref{lem:CasSysSta} with $\chi_1=\tilde \upsilon$ and $\chi_2=\tilde x$ together with Lemma \ref{lem:do}, the bound of $\tilde x$ is
\begin{equation}
		\| \tilde x(t)\| \leq \gamma_{\tilde x}(\|\tilde  x^{\rm{s}}(t_0)\|) \kappa^{\vartheta}(t-t_0)\exp(\tilde \alpha^{\rm s}(t-t_0)) \label{eq:tilde_x_bound}
	\end{equation}
	where $\tilde  x^{\rm{s}}=[\tilde \upsilon\t,\tilde x\t]\t$ and $\tilde \alpha^{\rm s}$ is some positive constant

We proceed to demonstrate the boundedness of $\phi_1(t)$ and $\phi_3(t)$ for $t\geq t_0$.
The proof for $\phi_1(t)$ closely follows that presented in Theorem~\ref{the1}.
	For $\phi_3(t)$,  it is sufficient to demonstrate that $\mu(t)\tilde\upsilon(t)<\infty$ and  $\mu(t)\tilde x(t)<\infty$ for $t\geq t_0$, by
	noting $\|\tilde \upsilon_i(t)\|\leq \|\tilde \upsilon(t)\|$, $\|\tilde x_i(t)\|\leq \|\tilde x(t)\|$, and $ y_i-C^\m_i  \hat{x}_i-D^\m_i  u_i-F^\m_i  \upsilon_i
	= -C^\m_i  \tilde{x}_i -F^\m_i \tilde \upsilon_i $.
	The proof for $\mu(t)\tilde\upsilon(t)$ is same as \eqref{mutermupsilon} in Theorem~\ref{the1}.
	By \eqref{eq:tilde_x_bound}, one has
	  \begin{equation}\label{mutermtildex}
\begin{aligned}
		\| \mu(t)\tilde x(t) \|
 \leq   \varepsilon_{\tilde x} \kappa^{\vartheta-1}(t-t_0)\exp(\tilde\alpha^{\rm s}(t-t_0))
\end{aligned}
	\end{equation}
for some positive finite constant $\varepsilon_{\tilde x}$.  The proof is thus completed.\eproof

The following theorem presents the results of the PTCOR implementation employing measurement output feedback
	\bthm\label{the2}
	Consider the closed-loop system composed of  the MASs (\ref{sys1}), the exosystem (\ref{sys2}), the distributed observer (\ref{18}), and the
	measurement feedback controller \eqref{hatx}-\eqref{40} under Assumptions~\ref{ass3}, \ref{ass4}, and \ref{ass2}.
	Suppose the parameters $\psi$, $\bar K_i$, $\tilde K_i$, and   $K_i$
	are selected as specified in Theorem~\ref{the1}, while the parameters $L_{i}$ and $\tilde L_{i}$
	are selected according to   Lemma~\ref{lem5},  with  additional
	conditions  $\vartheta_i\geq\theta_i+3/2$, $\psi \rho_H\geq \theta_i+\|\tilde L_{i}F_i^\m\|^2/2+1$ for  $i\in\mathcal{\bar V}$.
	Then, the PTCOR problem is solved in the sense that, for $i\in\bar{\mathcal V}$,
	the regulated output $e_i$ achieves prescribed-time convergence
	towards zero within $T+t_0$, and remains as zero after $T+t_0$.
	Moreover, the state of the closed-loop system and the $\mu(t)$-dependent terms $\phi_1(t)$, $\phi_3(t)$, and $\phi_4(t)$ in (\ref{eq:phi_signal})
	are bounded for all $t\geq t_0$.  \ethm
	
	\proofnow
Let $\chi_1=[\tilde \upsilon\t,\tilde x\t]\t$. Define Lyapunov function candidate as
\begin{equation*}
U(\chi_1)=\chi_1 \t P   \chi_1
\end{equation*}
 where $P=\mbox{diag}\{P_H,P_{L}\}$. Then, the time derivative along trajectory of (\ref{19}) and (\ref{eq:dot_til_x}) satisfies
\begin{equation}\label{dot_V_1}
\dot U(\chi_1)\leq -2(\theta+1)\mu U(\chi_1)+\tilde\varpi_{1}U(\chi_1)
\end{equation}
where $\tilde\varpi _{1}=(2\max\{\|A-L_{1}C^\m\|,\|P_H\otimes S_0\|\}+\|E-L_{1}F^\m\|)\lambda^{-1}_{\min}(P)$ and $\theta=\min\{\theta_1\cdots,\theta_N\}>1$.
For the   $\bar x$-dynamics in \eqref{dot_bar_x}, let Lyapunov function be $
	W(\bar x)={\bar x}\t P_{K}\bar x$. Then, $\dot W(\bar x)$ satisfies
 \begin{equation}\label{dot_V_2}
 \dot W(\bar x)\leq -2\theta\mu W(\bar x)+\tilde\varpi_{2} W(\bar x)+\tilde \varpi_{3}\mu^2\|\chi_1\|^2
 \end{equation}
 where
$\tilde\varpi_{2}=(2\|P_{K}\|\|A_{c}\|+4)\lambda^{-1}_{\min}(P_{K })$ and $\tilde\varpi_{3}=\max\{T^2\|B\bar K\|^2+\|BK\|^2,T^2\|B\tilde K\|^2+\|BKX\|^2\}$.

 The   regulated output $e$ in \eqref{e_2} satisfies
 \begin{align}
\|e(t)\|&\leq (T\|C_{c}\|+\|DK\|)\mu (t)\|\bar x(t)\|\notag\\
&\quad+(T\|D\tilde K\|+\|DKX\|)\mu(t)\|\tilde\upsilon (t)\|\notag\\
&\quad +(T\|D\bar K\|+\|DK\|)\|\tilde x\|.\label{eq:e_3}
\end{align}
Note that the dynamics \eqref{dot_V_1}, \eqref{dot_V_2}, and \eqref{eq:e_3} satisfy conditions in Lemma \ref{lem:CasSysSta} with  $\chi_2=\bar x$. Let {\blue $\chi(t_0)=[\chi_1(t_0)\t,\chi_2(t_0)\t]\t=[\tilde\upsilon(t_0)\t, \tilde x(t_0)\t, \bar x(t_0)\t]\t$}, by Lemma \ref{lem:CasSysSta}, $\bar x$  and $e$ satisfy
\begin{equation}\label{eq:e_x-1}
\begin{aligned}
\|\bar x(t)\|&\leq  \bar\gamma_{\chi }(\|\chi(t_0)\|)\kappa^{\theta}(t-t_0) \exp(\bar \alpha(t-t_0))\\
\|e(t)\|&\leq \bar\gamma_{e }(\|\chi(t_0)\|)\kappa^{\theta-1}(t-t_0) \exp(\bar \alpha(t-t_0))
\end{aligned}
\end{equation}
for some $\bar\gamma_{\chi }$, $\bar \gamma_{e}\in \mathcal K$ and some positive finite constant $\bar \alpha$.
	The PTCOR problem is thus solved.
{\blue By \eqref{40} and \eqref{StaTrans}, $\tilde u$ can be expressed as
\begin{align}
\tilde u = \bar K(\bar x+\tilde x) + \tilde K \tilde \upsilon + \mu (\bar x +\tilde x -X\tilde \upsilon).\label{eq:bound-tilde_u-1}
\end{align}
Then, according to \eqref{eq:e_x-1}, the bound of $\tilde u$ is
\begin{equation}
\|\tilde u(t)\|\leq \bar\gamma_{\tilde u }(\|\chi(t_0)\|)\kappa^{\theta-1}(t-t_0) \exp(\bar \alpha(t-t_0))\notag
\end{equation}
where $\bar \gamma_{\tilde u}\in \mathcal K$.}
Hence, the states $v_i$, $\hat x_i$ and $x_i$, and controller $u_i$  of the closed-loop system
	are bounded   for all $t\geq t_0$ due to  \eqref{eq:e_x-1}, \eqref{eq:tilde_x_bound} and \eqref{eq:bound-tilde_u-1}.
	
	It has been proved in  Lemma~\ref{lem5} that the $\mu(t)$-dependent terms $\phi_1(t)$ and $\phi_3(t)$ are bounded for  $t\geq t_0$.
	What is left is to prove that  $\phi_4(t)$ is bounded for all $t\geq t_0$.
	Noting $\hat{x}_i-X_{i}\upsilon_i =\bar x_i -X_{i}\tilde \upsilon_i +  \tilde x_i $,
	it suffices to show that $\mu(t)\tilde\upsilon(t)$, $\mu(t)\bar x(t)$, and $\mu(t)\tilde x(t)$ are bounded,
	which is indeed true due to \eqref{mutermupsilon}  and  \eqref{mutermtildex}.
	\eproof

{\blue
 \brem
Due to the cascaded structure of the dynamics of $\tilde{\upsilon}$, $\tilde{x}$, and $\bar{x}$,  the prescribed-time stabilization requires more than the conditions that  the feedback gain $\psi$ is positive and  the closed-loop matrices $A_{ci}$ and $A_{Li}$ are Hurwitz, as commonly assumed in COR. More conditions must be imposed. For the state feedback, the feedback gains $\psi$ and $K_i$ must satisfy the conditions specified in \eqref{eq:theta_i} and \eqref{eq:theta_i}. For the output measurement feedback, the feedback gains $\psi$ and $\tilde{L}_{i}$ must fulfill the requirements in \eqref{eq:vartheta_i} and \eqref{eq:vartheta_i}, thus making the design of local state observers distinct from the existing results found in \cite{holloway2019prescribed66, holloway2019prescribed78}.
\erem
}

{\blue
 \brem
Consider a cascaded system described by $\dot \xi = f(\xi,t)$ and $\dot z = h(z,t) + \psi(z,\xi,t)$, where $\xi\in \mathbb R^n$ and $z\in \mathbb R^s$. 
Note that the  term $\psi(z,\xi,t)$ may induce finite-escape time for $z$-subsystem   when $\xi(t)$ is not appropriately controlled \cite{sepulchre2002trading98}, even if
 $z=0$ is  a stable equilibrium  of the system $\dot z = h(z,t)$. 
 
In this paper, we demonstrate that the singularity of the solution caused by the piecewise continuous function $\mu(t)$ can be addressed by the generalized Filippov solution  proposed in \cite{SEEBER91}. Moreover,
the finite-escape time issue can be resolved through the state feedback in the cascaded system as described by \eqref{19} and \eqref{domegai1}, as well as  the  measurement output feedback for the cascaded system outlined in \eqref{19}, \eqref{dot_til_x}, and \eqref{dot_bar_x}. The design of the $\tilde \upsilon$-dynamics in \eqref{19}, which admits a  PTLF as  in \eqref{eq:V_1} and meets the condition in \eqref{eq:eta}, ensures that $\tilde \upsilon$ converges with the desired prescribed-time convergence rate. This prevents $\bar x$ and $\tilde x$ from diverging as the time approaches $T+t_0$, thereby avoiding finite-time escape.
\erem
}
{\blue
\brem
The implementation of PTCOR requires more stringent condition that matrices $B_i$ and $C_i^\m$  must satisfy Assumption \ref{ass2},  while  the condition of the asymptotic convergence for the COR  is that $(A_i,B_i)$ is stabilizable and $(C_i^\m, A_i)$ is detectable. 
\erem}

\section{Simulation}\label{Sec:Simulation}
In this section, we verify the proposed PTCOR algorithm by two numerical simulations.
\subsection{Numerical Example 1}
Consider the  MASs of RLC circuits,  each of which is shown in Fig.~\ref{fig6}. Let $x=[u_c,i_L]\t$ be the system state variables. According to
the Kirchhoff laws, we have the following equations
\begin{equation}\label{eq29}
\begin{aligned}
	u_c+R_2C\dot{u}_c-L\dot{i}_L&=u_2 \\
	R_1i_L+R_1C\dot{u}_c+L\dot{i}_L&=u_1.
\end{aligned}
\end{equation}
\begin{figure}[H]
	\centering
	\begin{minipage}{0.49\linewidth}
		\centering
		\includegraphics[width=1\linewidth]{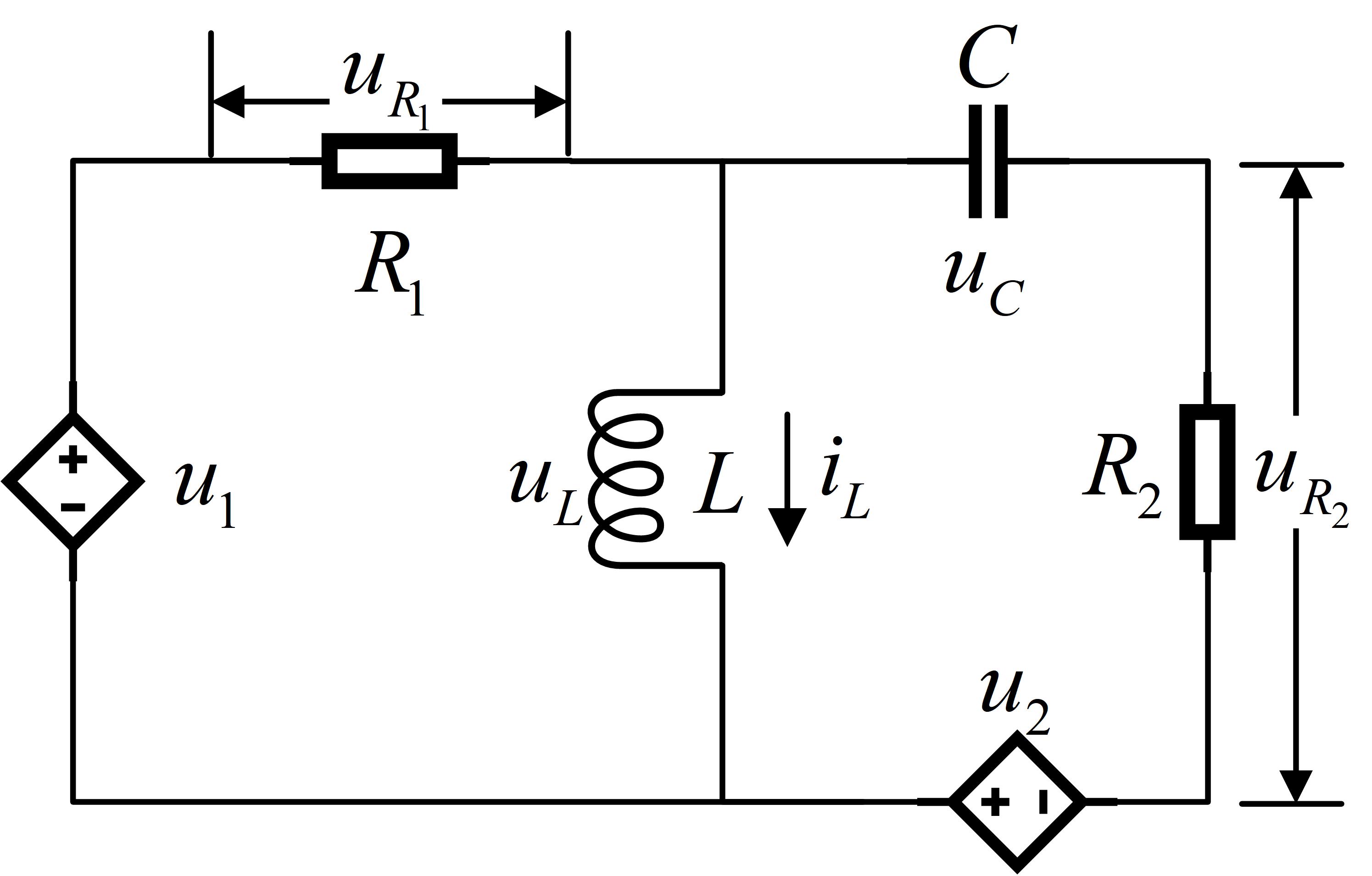}\vspace{5mm}
\caption{The structure of a circuit system. }
		\label{fig6}
	\end{minipage}
	\begin{minipage}{0.49\linewidth}
		\centering
		\includegraphics[width=1\linewidth]{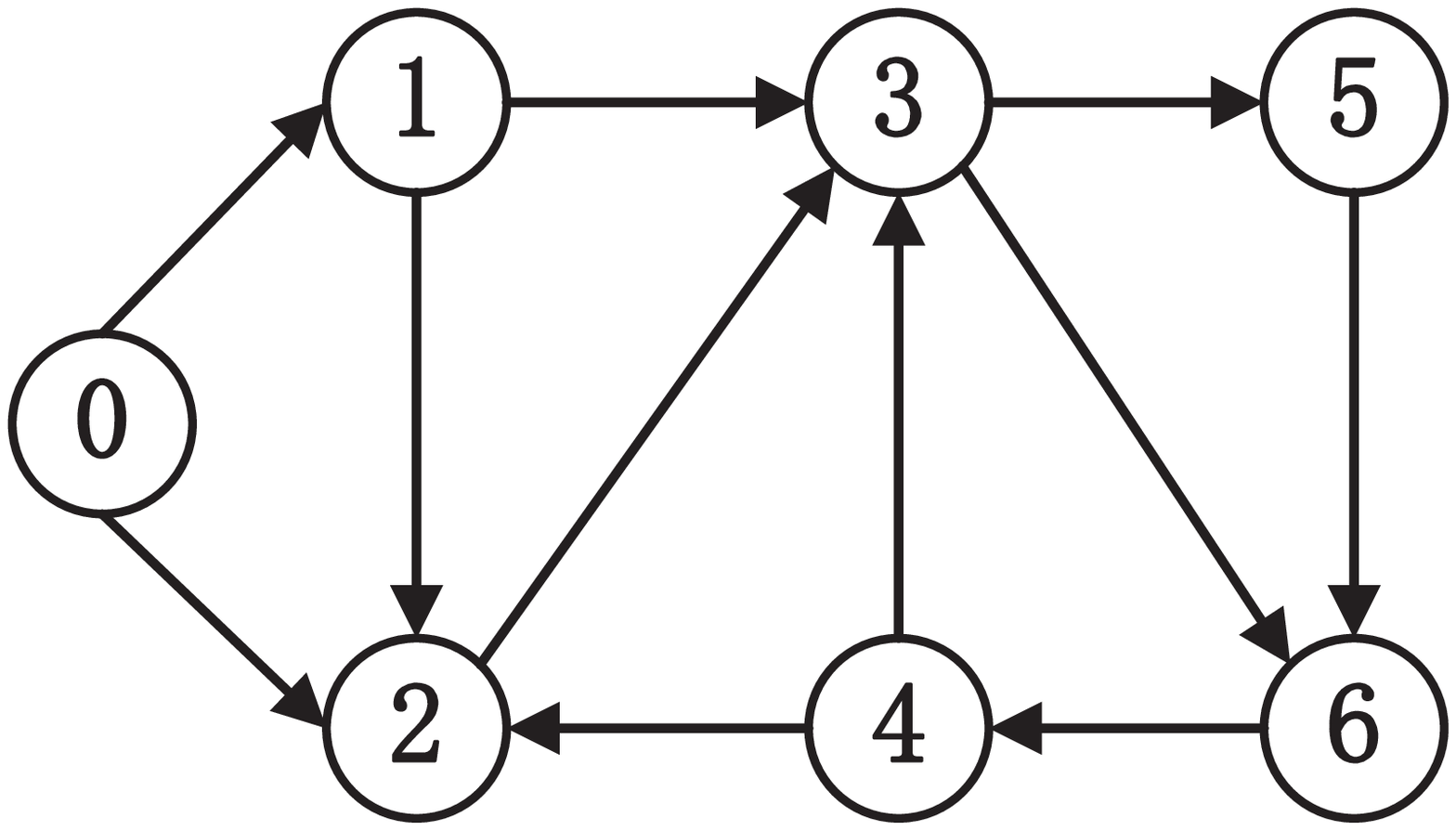}\vspace{2mm}
		\caption{The communication graph $\mathcal{G}$.}
		\label{fig1}
	\end{minipage}
\end{figure}
%
%
 Let $u=[u_1,u_2]\t$ denote the system control input, $e=[u_{R1}, u_{R2}]\t-[u_{d1},u_{d2}]\t$ the system output, $y=[u_C,u_L]\t$ the system measurement output, and $\upsilon_0=[u_{d1},u_{d2}]\t$ the reference input  generated by
  \begin{equation}\label{eq30}
	\dot{\upsilon}_0=\left[\begin{array}{cc}0&1\\-1&0\end{array}\right]\upsilon_0,\quad u_d(t_0)=\left[\begin{array}{c}1\\ 1\end{array}\right]
\end{equation}
where we note $\upsilon_0(t)$ is two sinusoidal functions.

\begin{figure}[!htp]
	\centering
	\includegraphics[width=0.9\linewidth]{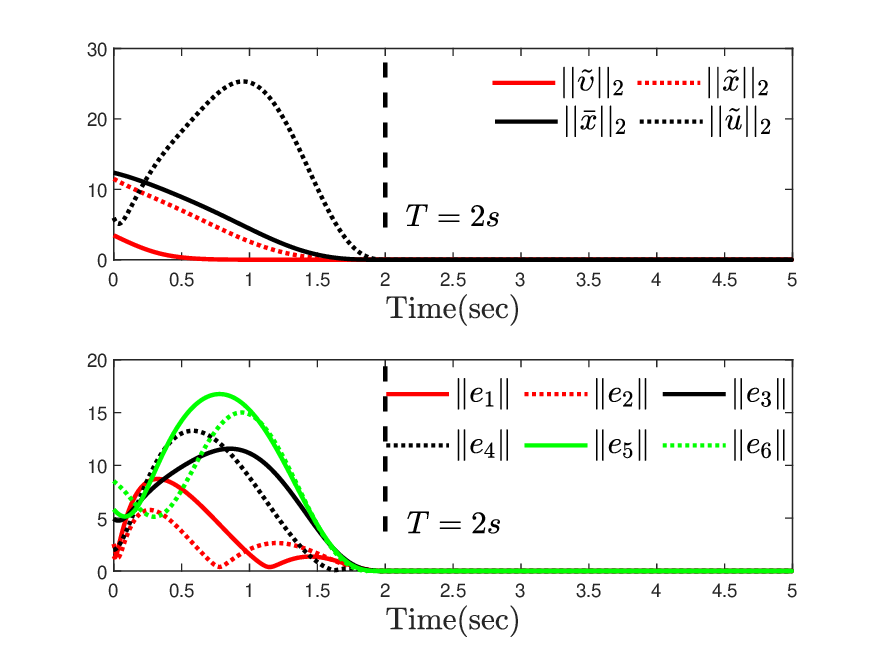}\\

	\caption{ {\blue Trajectories of $\Vert \tilde \upsilon\Vert_2$, $\Vert \tilde x\Vert_2$, $\Vert \bar x\Vert_2$, $\Vert\tilde u\Vert_2$ and local regulated output tracking errors $e_i,i=1,\cdots,6$.}}
	\label{fig2}
\end{figure}

The leader of the MASs is governed by  (\ref{eq30}) and the six followers by  (\ref{eq29}). The communication graph is shown in Fig. \ref{fig1}. Let $\bar{R}=1/(R_1+R_2)$, the state space equation of followers can be expressed as form (\ref{sys1}) with $A_i=\bar R[-1/C,-R_1/C;R_1/L,-R_1R_2/L]$,
$B_i=\bar R[{1}/{C},{1}/{C};{R_2}/{L},-{R_1}/{L}]$,
 $E_i=0$, $C_i=\bar R[-R_1,R_1R_2;-R_2, -R_1R_2]$, $D_i=\bar R[R_1,R_1;R_2,R_2]$, $F_i=I$,
 $C_i^\m=\bar R[-1,-R_1;R_1,-R_1R_2]$, $D_i^\m =\bar R[1,1;R_1,-R_1]$, and
$F_i^\m=0,i=1,\cdots,6$.
The circuit parameters  are chosen as $R_1=3\Omega$, $R_2=1\Omega$, $C=1$F, and $L=1$H. The initial conditions of $x_i$ are chosen as  $x_1(0)=[2,2]\t$, $x_2(0)=[0,2]\t$, $x_3(0)=[2,-4]\t$, $x_4(0)=[4,0]\t$, $x_5(0)=[4,-4]\t$, and  $x_6(0)=[-6,4]\t$, and the initial conditions of the distributed and local observers are $\upsilon_i=0$ and  $\hat{x}_i(0)=0,i=1,\cdots,6$. Let the initial time $t_0=0$, the prescribed-time $T=2s$, and total simulation time is $5s$. The control parameters are chosen for the measurement output feedback controller according to Theorem \ref{the2} as $\psi=8$, $\bar K_i =[0,0;0,0]$, $\tilde K_i=[-2,-0.33;0,-0.67]$, $K_i=[-9,-3;-3,3]$, $L_{i}=[1,-2;2,-0.3]$, and $\tilde L_{i}=[-4,4;-4,-1.33]$, $i=1,\cdots,6$. The simulation results depicted in Fig.~\ref{fig2} reveal that all internal signals remain bounded for $t\geq t_0$, with $e_i(t)$ achieving prescribed-time convergence towards zero.

\begin{figure}[!htp]
	\centering \includegraphics[width=0.9\linewidth]{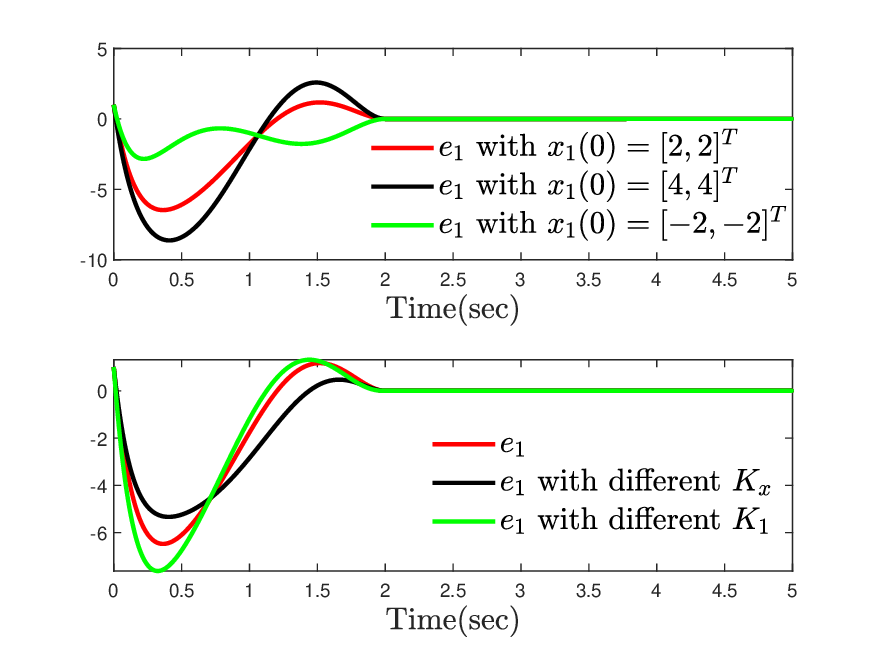}\\
\vspace{2mm}
	\caption{Trajectories of $e_1$ with different initial conditions and control parameters.}
	\label{fig4}
\end{figure}

Furthermore, we replicate the simulations using varied initial values while maintaining the same set of control parameters, and vice versa, altering the control parameters while retaining the same initial values. It is observed that the prescribed-time convergence is always guaranteed.
For example, the convergence  of the regulated output $e_1$ is plotted in Fig.~\ref{fig4} to demonstrate the regulation performance.
The plots illustrate that the convergence time of $e_{11}$ remains unaffected by both the initial values and the control parameters, instead being solely determined by the prescribed value of $T=2s$.
\subsection{Numerical Example 2}
\begin{figure}[!htp]
	\centering \includegraphics[width=0.9\linewidth]{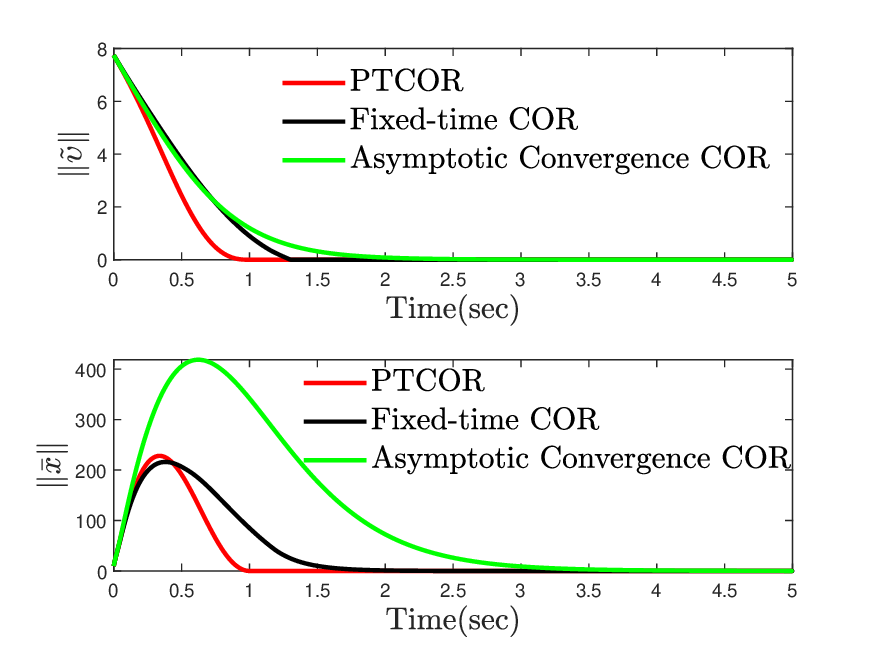}\\
\vspace{2mm}
	\caption{Trajectories of $\|\tilde \upsilon \|$ and $\| \bar x\|$ under different COR algorithms. }
	\label{fig5}
\end{figure}

{\blue In this subsection, we consider the voltage control problem for CCVSIs. For simplicity, the connection of the microgrid system is simplified as in Fig. \ref{fig1}. According to \cite{Caihe52}, the voltage control problem for CCVSIs under the graph  in Fig. \ref{fig1} can be converted into the COR problem  of  linear MASs in \eqref{sys1} with $A_i = [-b_{1i}, \omega_i; -\omega_i, -b_{1i}]$, $B_i = \diag \{b_{2i} , b_{2i}\}$, $E_i =  [-b_{2i},0,0,0;0,-b_{2i},0,0]$, $C_i = I_2$, $D_i = 0_{2\times 2}$, $F_i = [0,0,-1,0; 0,0,0,-1]$, $C_i^\m = C_i$, $D_i^\m = D_i$, and $F_i^\m =F_i, i=1,\cdots,6$, where $b_{1i} = R_{fi}/L_{fi}$, $b_{2i} = 1/ L_{fi}$, and $\omega_i$ is the frequency of the reference frame of $i$-th CCVSIs.
The system matrix for leader system \eqref{sys2} is $S_0 =  [0,k_\omega (\omega^* -\bar \omega),0,0;  k_\varsigma (\varsigma^*  -\bar \varsigma), 0,0,0; 0,0,0,0;0,0,0,0]$ and the initial value is $\upsilon_0 (t_0) = [0,0,i^*_{od}, i^*_{oq} ]$, where $k_\varpi $ and $k_\varsigma$ are the integral gains, $\bar \omega$ and $\bar \varsigma$ are the average frequency and voltage magnitude of the microgrid, respectively, $\omega^*$ and $\varsigma^*$ are the nominal frequency and voltage of the microgrid, $i^*_{od}$ and $i^*_{oq}$ are the optimal output currents for the CCVSIs.

The parameters of controlled MASs are $R_{fi} = 0.1\Omega$, $L_{fi} = 0.00135\rm{H}$, $\omega_i = 50\rm{Hz}$ for $i=1,\cdots,6$, $\bar\omega = 48.5\rm{Hz}$, $\bar\varsigma = 375\rm{v}$, $\omega^\star = 50\rm{Hz}$, $\varsigma^\star = 380\rm{V}$, $k_\omega = 0.5$, $k_\varsigma = 0.5$, $i_{od}^\star = 3$ and $i_{oq}^\star = -1$. The initial conditions for state, distributed observers and local state observers are $x_i(t_0) = [3,3]\t$, $\upsilon_i(t_0) = [1,1,1,1]\t$ and $\hat x_i (t_0) = [1,1]\t$ for $i=1,\cdots,6$. Let the initial time $t_0 = 0$, the prescribed-time $T=1s$, and total simulation time is $5s$. The control parameters   are chosen for the measurement output feedback controller according to Theorem \ref{the2} as $\psi =4$, $\bar K_i = [0.0973, -0.0675; 0.0675, 0.0973]$, $\tilde K_i = [1, 0, 0.0027, 0; 0, 1, 0, 0.0027]$, $L_i = [0, 50; -50,0]$, $\tilde L_i = [-1,0;0,-1]$ for $i = 1,\cdots,6$.

To verify the advantages of our proposed PTCOR algorithm, we conduct the comparison simulations with fixed-time COR algorithm in \cite{song2021distributed103} and asymptotic convergence COR in \cite{huangjie2}. The fixed-time COR algorithm is designed as $\dot \upsilon_i = S_0 \upsilon_i + c_1 \chi_i + c_2 \mbox{sign} (\chi_i) + c_3 \mbox{sig}(\chi_i)^{c_4}$, $\dot {\hat x}_i = A_i \hat x_i + B_i u_i + E_i \upsilon_i + L_i \tilde \chi_i + \tilde L_i \mbox{sign} (\tilde \chi _i) + \tilde L_i \mbox{sig}(\tilde \chi_i)^{c_4}$, and $u_i = \bar K_i \hat x_i + \tilde K_i \upsilon_i + K_i \mbox{sign}(\hat x_i - \upsilon_i) + K_i \mbox{sig}(\hat x_i -\upsilon_i)^{c_4}$,
where
$c_1 = 5$, $c_2 =5$, $c_3 = 5$, $c_4 = 1.1$, $\chi_i = \sum_{j=0}^N a_{ij}(\upsilon_j-\upsilon_i)$, $\tilde \chi _i = y_i -C_i^\m \hat x_i - D_i^\m u_i -F_i ^\m \upsilon_i$,
$\mbox{sign}(\chi_i) = [\mbox{sign}(\chi_{i1}),\cdots,\mbox{sign}(\chi_{iq})]\t$ its element-wise sign function vector, $\mbox{sig}(\chi^{c_4}) = [\mbox{sign}(\chi_{i1})|\chi_{i1}|^{c_4}, \cdots, \mbox{sign}(\chi_{iq})|\chi_{iq}|^{c_4}]\t$, and the matrices $L_i$, $\tilde L_i$, $\bar K_i$, $\tilde K_i$ and $K_i$ are same with the PTCOR algorithm. The asymptotic convergence COR algorithm is designed as $\dot \upsilon_i = S_0 \upsilon_i + \psi \chi_i$, $\hat x_i = A_i \hat x_i + B_i u_i + E_i \upsilon_i + L_i \tilde \chi_i $, and $u_i= \bar K_i \hat x_i + \tilde K_i \upsilon_i$.
The simulation results are presented in Fig. \ref{fig5} - Fig. \ref{fig7}, which show that the convergence performance of our proposed PTCOR algorithm is better. 
}

\begin{figure}[!htp]
	\centering \includegraphics[width=0.9\linewidth]{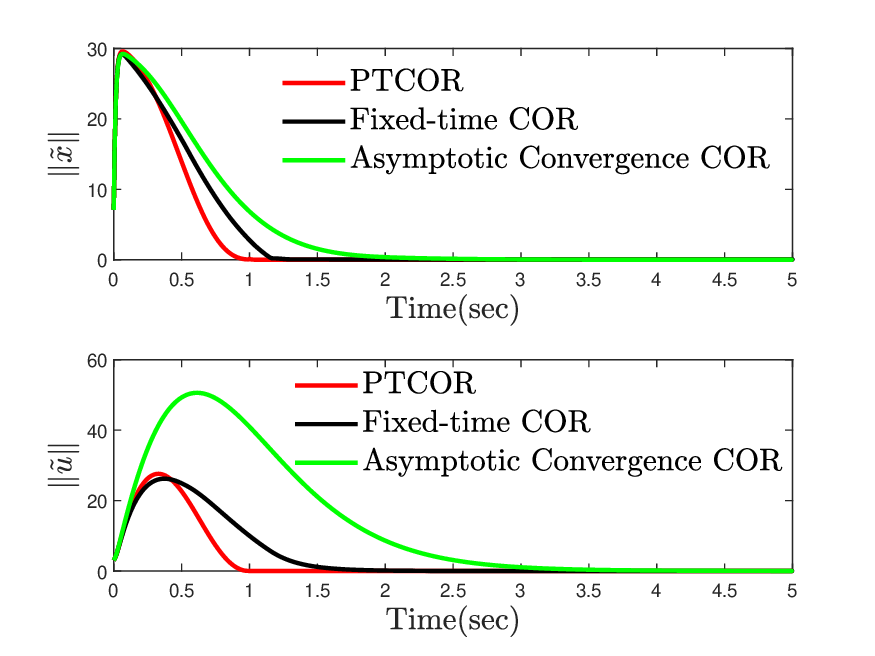}\\
\vspace{2mm}
	\caption{Trajectories of $\|\tilde x \|$ and $\| \tilde u \|$ under different COR algorithms. }
	\label{fig6}
\end{figure}

\begin{figure}[!htp]
	\centering \includegraphics[width=0.9\linewidth]{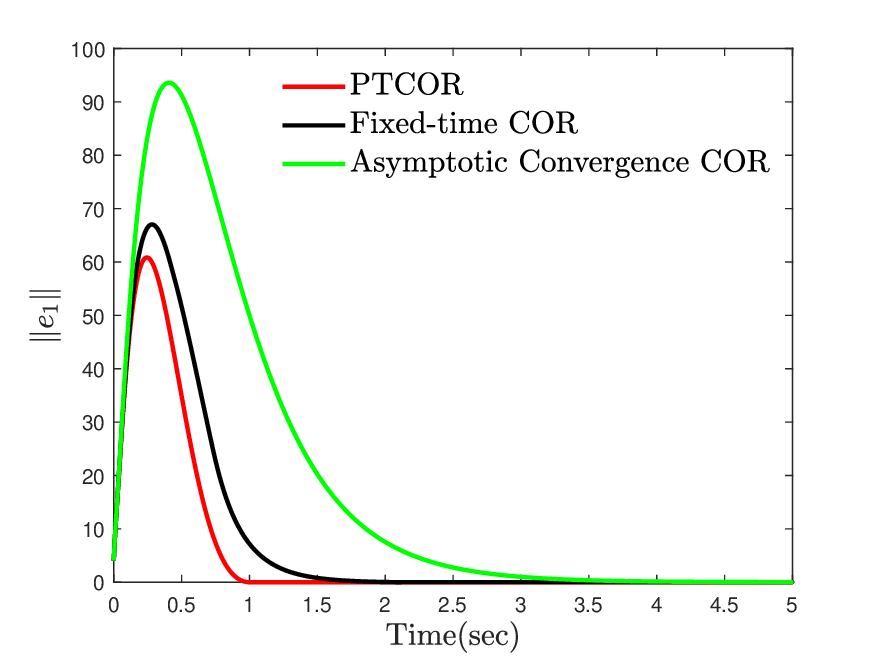}\\
\vspace{2mm}
	\caption{Trajectories of $\|e_1\|$ under different COR algorithms. }
	\label{fig7}
\end{figure}

\section{Conclusion}\label{sec:conclusion}
In this paper, we focus on tackling the PRCOR problem for linear  heterogeneous MASs. Our proposed control approach stands out for its capability to attain COR within a prescribed-time duration $T$, irrespective of initial conditions or other design parameters. Moreover, all internal signals in the closed-loop system are proved to be bounded.  Extending the proposed methodology to investigate PTCOR for discrete-time MASs would be our further research direction.

\bibliographystyle{ieeetr}        
\bibliography{ref}

\begin{IEEEbiography}[{\includegraphics[width=1in, height=1.25in, clip, keepaspectratio]{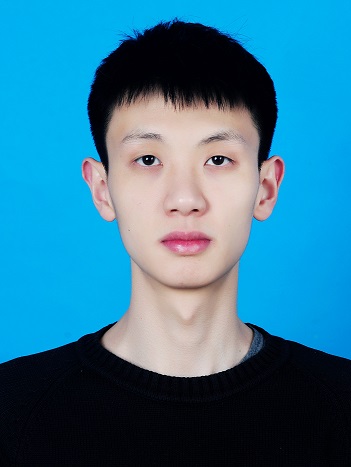}}]{Gewei Zuo}
	received the B.E. degree in automation from Xian University of Architecture and Technology, Xian, Shanxi, China, in 2019, and the M. E. Degree in control theory and engineering from Chongqing University, Chongqing, China, in 2022. He is currently pursuing the Ph.D. degree in control science and engineering with the school of Artificial Intelligence and Automation with Huazhong University of Science and Technology, Wuhan, Hubei, China.
	His research interests include Nonlinear System Control Theory, Distributed Cooperative Control and Distributed Convex Optimization.
\end{IEEEbiography}

\begin{IEEEbiography}[{\includegraphics[width=1in, height=1.25in, clip, keepaspectratio]{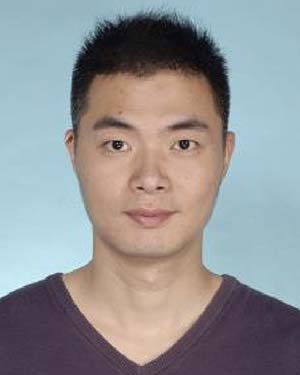}}]{Lijun Zhu}
	 received the Ph.D. degree in Electrical Engineering from University of Newcastle, Callaghan, Australia, in 2013. He is now a Professor in the School of Artificial Intelligence and Automation, Huazhong University of Science and Technology, Wuhan, China. Prior to this, he was a post-doctoral Fellow at the University of Hong Kong and the University of New- castle. His research interests include power systems, multi-agent systems and nonlinear systems analysis and control.
	\end{IEEEbiography}

\begin{IEEEbiography}[{\includegraphics[width=1in, height=1.25in, clip, keepaspectratio]{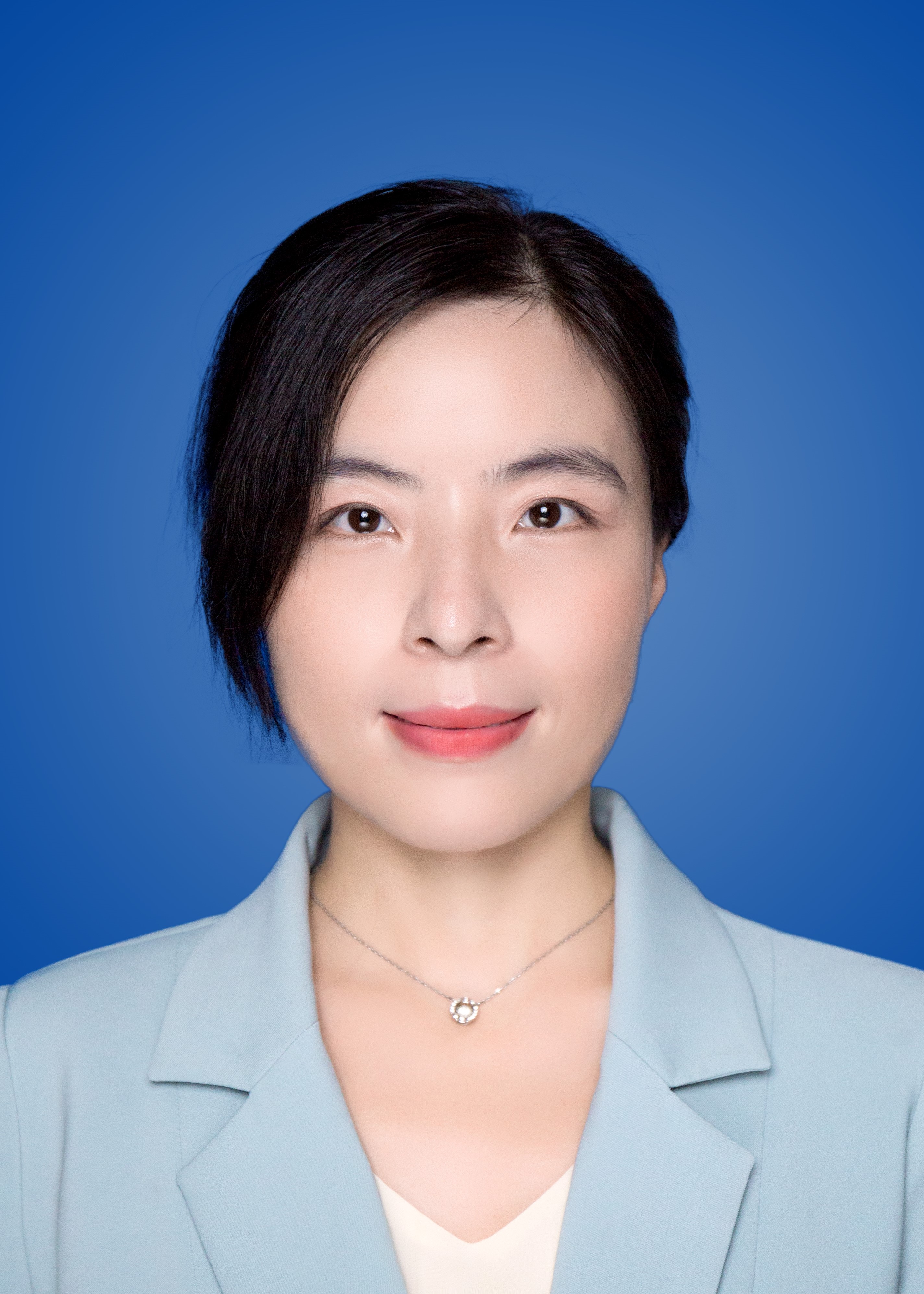}}]{Yujuan Wang}
	 received the Ph.D. degree in the School of Automation, Chongqing University, Chongqing, China, in 2016. She is now a Professor in the School of Automation, Chongqing University, Chongqing, China. Prior to this, she was a post-doctoral Fellow at the University of Hong Kong and a Joint Ph.D. Student at University of Texas at Arlington. Her research interests include nonlinear system control, distributed control, cooperative
adaptive control, fault-tolerant control.
	\end{IEEEbiography}

\begin{IEEEbiography}[{\includegraphics[width=1in, height=1.25in, clip, keepaspectratio]{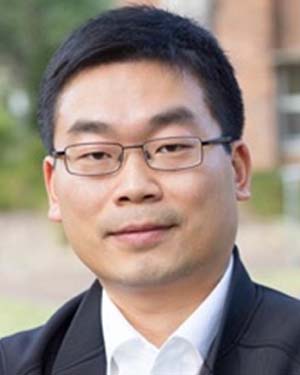}}]{Zhiyong Chen} (Senior Member, IEEE) received the B.E. degree in automation from the University
of Science and Technology of China, Hefei, China, in 2000 and the M.Phil. and Ph.D. degrees in mechanical and automation engineering from the Chinese University of Hong Kong, Hong Kong, in 2002 and 2005, respectively.
He was a Research Associate with the University of Virginia, Charlottesville, VA, USA, from
2005 to 2006. In 2006, he joined the University of Newcastle, Callaghan, NSW, Australia, where
he is currently a Professor. He was also a Changjiang Chair Professor with Central South University, Changsha, China. His research interests include nonlinear systems and control, biological systems, and reinforcement
learning.

Dr. Chen is/was an Associate Editor for \emph {Automatica, IEEE Transactions
on Automatic Control, IEEE Transactions on Neural
Networks and Learning Systems, and IEEE Transactions on Cybernetics}.
	\end{IEEEbiography}
\end{document}